\documentclass{lmcs}
\pdfoutput=1
\usepackage[utf8]{inputenc}

\usepackage{lastpage}
\lmcsdoi{20}{3}{25}
\lmcsheading{}{\pageref{LastPage}}{}{}%
{Dec.~18,~2023}{Sep.~18,~2024}{}


\input{imports}
\newcommand{\ex}[1]{\textsf{#1}}

\newcommand{\nook}{\ex{Nook}}
\newcommand{\house}{\ex{House}}
\newcommand{\houses}{\ex{Houses}}

\newcommand{\architect}{\ex{architect}}
\newcommand{\houseType}{\ex{HouseType}}
\newcommand{\cube}{\ex{Cube}}
\newcommand{\cubes}{\ex{Cubes}}
\newcommand{\villa}{\ex{Villa}}

\newcommand{\type}{\ex{type}}

\newcommand{\plans}{\ex{Plans}}
\newcommand{\plan}{\ex{Plan}}
\newcommand{\construction}{\ex{Construction}}
\newcommand{\constructions}{\ex{Constructions}}
\newcommand{\constructionStep}{\ex{Construction Step}}
\newcommand{\constructionSteps}{\ex{Construction Steps}}
\newcommand{\cellar}{\ex{Cellar}}
\newcommand{\floor}{\ex{Floor}}
\newcommand{\floors}{\ex{Floors}}
\newcommand{\saddleRoof}{\ex{Saddle Roof}}
\newcommand{\company}{\ex{company}}

\newcommand{\nookRule}{\ex{Nook Rule}}
\newcommand{\cubeRule}{\ex{Cube Rule}}
\newcommand{\cubeRules}{\ex{Cube Rules}}
\newcommand{\villaRule}{\ex{Villa Rule}}
\newcommand{\doubleCubeRule}{\ex{Double Cube Rule}}

\newcommand{\cubeFWDRule}{\ex{Cube Forward Rule}}

\newcommand{\doubleCubeFWDRule}{\ex{Double Cube Forward Rule}}

\newcommand{\shortcutRule}{\textit{Short-Cut Rule}}
\newcommand{\shortcutRules}{\textit{Short-Cut Rules}}

\newcommand*\circled[1]{\tikz[baseline=(char.base)]{%
		\node[shape=circle,fill=white!20,draw,inner sep=2pt] (char) {#1};}}

\newcommand{\marked}{$\text{\rlap{$\checkmark$}}\square$}
\newcommand{\marking}{$\square \rightarrow $ \marked}

\definecolor{orange}{RGB}{255,145,0}

\begin{document}
\title[Advanced Model Consistency Restoration]{Advanced Model Consistency Restoration with Higher-Order Short-Cut Rules}
\thanks{This work was partially funded by the German Research Foundation
(DFG), project “Triple Graph Grammars (TGG) 3.0”.}

\author[L.~Fritsche]{Lars Fritsche\lmcsorcid{0000-0003-4996-4639}}[a]
\author[J.~Kosiol]{Jens Kosiol\lmcsorcid{0000-0003-4733-2777}}[b,c]
\author[A.~Lauer]{Alexander Lauer\lmcsorcid{0009-0001-9077-9817}}[b]
\author[A.~Möller]{Adrian Möller\lmcsorcid{0009-0002-2871-6177}}[a]
\author[A.~Schürr]{Andy Schürr\lmcsorcid{0000-0001-8100-1109}}[a]

\address{Technical University Darmstadt, Darmstadt, Germany}
\email{\{lars.fritsche, andy.schuerr\}@es.tu-darmstadt.de, adrian.moeller@stud.tu-darmstadt.de}

\address{Philipps-Universität Marburg, Marburg, Germany}
\email{kosiolje@mathematik.uni-marburg.de, alexander.lauer@uni-marburg.de}

\address{Universität Kassel, Kassel, Germany}

\begin{abstract}
Sequential model synchronisation is the task of propagating changes from one model to another correlated one to restore consistency. 
It is challenging to perform this propagation in a least-changing way that avoids unnecessary deletions (which might cause information loss). 
From a theoretical point of view, so-called \emph{short-cut (SC) rules} have been developed that enable provably correct propagation of changes while avoiding information loss. 
However, to be able to react to every possible change, an infinite set of such rules might be necessary. 
Practically, only small sets of pre-computed \emph{basic SC rules} have been used, severely restricting the kind of changes that can be propagated without loss of information. 
In this work, we close that gap by developing an approach to compute more complex required SC rules on-the-fly during synchronisation. 
These \emph{higher-order SC rules} allow us to cope with more complex scenarios when multiple changes must be handled in one step. 
We implemented our approach in the model transformation tool eMoflon. 
An evaluation shows that the overhead of computing \emph{higher-order SC rules} on-the-fly is tolerable and at times even improves the overall performance. 
Above that, completely new scenarios can be dealt with without the loss of information.
\end{abstract}

\maketitle              % typeset the header of the contribution

\section{Introduction}
\label{chapter:introduction}
Model-Driven Engineering (MDE)~\cite{BCW17} provides the necessary means to tackle the challenges of modern software systems, which become more and more complex and distributed.
Using MDE, a system can be described by various models that describe different aspects and provide specific views onto the system itself, where each view may overlap to some degree with other views of the same system.
Consequently, if one view changes, we have to propagate these changes to other views that share the same information. 
This is necessary to ensure that all models together consistently represent the overall system state. 
The propagation process is often called \textit{model synchronisation}, where we distinguish between \textit{sequential} and \textit{concurrent} model synchronisation.
In this paper, we will focus on the former, where only one model is changed at a time, while the latter case would also incorporate changes to multiple models at once. 

As our methodology of choice, we employ \emph{Triple Graph Grammars} (TGGs)~\cite{Schurr94}, which are a declarative way to specify a consistency relationship between possibly heterogeneous models by means of a set of graph grammar rules.
By transforming these rules, we can automatically derive different \emph{consistency restoring operators} that served as central ingredients for various model management processes, such as batch translators~\cite{Schurr94, EhrigEEHT07, HermannEGO14}, consistency checkers~\cite{Leblebici18} or sequential~\cite{GieseW09, LauderAVS12, LeblebiciAFVS17, FritscheKST21} and concurrent synchronisers~\cite{OrejasPN20, FritscheKMST20, WeidmannFA20}. 
In general, synchronisers are required to avoid information loss. 
Most TGG-based approaches, in fact, do not meet this requirement: 
They propagate changes by deleting (parts of) the model that is to be synchronised and then retranslating (the missing parts) from the updated model~\cite{GieseW09, LauderAVS12, Hermann2015, Leblebici18, LeblebiciAFVS17}. 
This procedure is both inefficient and prone to loss of information that is private to the synchronised model. 
For this purpose, we introduced \textit{Short-Cut Rules}~\cite{FritscheKST18, FritscheKST21, KosiolT23}, which implement consistency- and information-preserving modifications. 
Compared to former approaches, \shortcutRules{} reduce (or even completely avoid) the unnecessary deletion of elements in the model that is to be synchronised but are able to reuse these elements and integrate them consistently.
In this way, deleting elements just to re-create them is avoided and information is preserved instead.
Even more, we can provide conditions under which the application of a \shortcutRule{} reestablishes or at least improves consistency~\cite{Kosiol22, KosiolT23}.

The translation and, thus, revocation of former translation steps is based on basic TGG rules that describe atomic consistency-preserving model changes. 
The same holds for our first works on \shortcutRules{}~\cite{FritscheKST18, FritscheKST21}, where we created them by combining pairs of basic TGG rules.
While the resulting \shortcutRules{} describe useful repairs in many situations, we found that there are cases where we have to repair multiple inconsistencies at once to yield a better result.
Hence, we here propose to concatenate TGG rules that describe a sequence of atomic model changes and derive \shortcutRules{} from these to repair multiple TGG rule applications at once.
Since there are usually infinitely many possibilities of concatenating TGG rules and thus infinitely many \shortcutRules{}, we propose an analysis that determines at runtime which TGG rules to concatenate to construct a \shortcutRule{} for a particular scenario. 
This allows us to also reduce information loss during synchronisation in scenarios where the \shortcutRules{} we computed so far did not suffice. 

In more details, we make the following contributions in this paper:
\begin{itemize}
    \item We introduce a runtime analysis for the derivation of what we call \textit{higher-order} \shortcutRules{}; their use promises to reduce the amount of lost information during model synchronisation. 
    \item We show how our proposed runtime analysis can be presented as an integer linear programming (ILP) problem~\cite{luenbergerLinear2016}. 
    \item We integrate our approach into eMoflon~\cite{WeidmannFA20}, a state-of-the-art graph transformation tool that implements TGGs.
    \item We evaluate our approach on a synthetic example; our key findings are that we can indeed preserve more information in certain cases and that while our implementation introduces a noticeable offset in some scenarios, in others we could even improve the runtime performance compared to our previous works.
\end{itemize}

This paper extends our previous contribution~\cite{FritscheKMS23}, where we were only able to convey the general concept behind finding \shortcutRules{} that repair multiple steps at once.
In this extension, we provide a detailed description of the construction process of \shortcutRules{} by presenting it as an ILP problem and, as promised in our previous work, evaluate our approach for another scenario that is particularly challenging for it.

This paper is structured as follows: 
We introduce TGGs and \shortcutRules{} together with our running example in \cref{chapter:fundamentals}.
In \cref{chapter:contribution}, we will present our novel analysis to determine non-trivial consistency-restoring operators in the form of \shortcutRules{} at runtime.
The construction process of \shortcutRules{} using optimisation techniques is presented in \Cref{chapter:contribution_extended}.
\Cref{chapter:evaluation} presents our evaluation investigating how much information is preserved and whether it comes with additional costs.
Finally, in \cref{chapter:relatedWork}, we discuss related works and summarise open challenges in \cref{chapter:conclusion}.

\section{Fundamentals}
\label{chapter:fundamentals}
This section provides an informal introduction to the state-of-the-art of TGG-based\linebreak information-preserving model synchronisation.
We begin with introducing TGGs along with our running example.
Then, we illustrate the need for synchronisation and exemplify \emph{precedence graphs}, which are the central data structure we use to keep track of inconsistencies between models and to direct our synchronisation process.
Finally, based on a small yet non-trivial synchronisation example, we will motivate the need for specific repair rules, namely \shortcutRules{}.

\subsection{Running Example \& Triple Graph Grammars}
\begin{figure}
	\centering
	\includegraphics[width=0.8\textwidth]{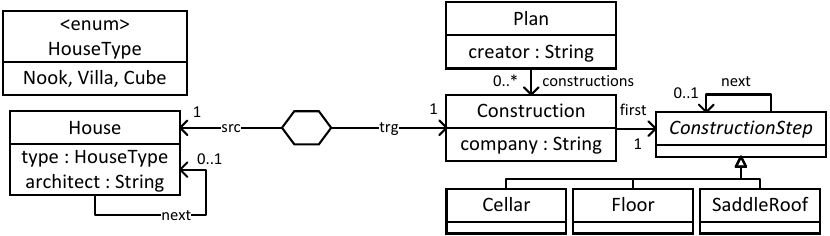}
	\caption{Source, Target and Correspondence Metamodel} 
	\label{fig:fundamentals:metamodel}
\end{figure}

In our running example throughout this paper, we will define consistency between terrace house and construction planning models to which we will refer in the following as source and target, respectively.
Due to the complexity of the derived repair operations later in this paper, we had to choose a rather small example.
Yet, TGGs have been successfully applied in industrial applications~\cite{Giese2012, Becker2003, Anjorin2020}.
As a first step, we introduce a third correspondence model between source and target, which connects elements from both sides and thus makes corresponding information traceable.
\Cref{fig:fundamentals:metamodel} depicts the three metamodels with the source metamodel on the left, the target metamodel on the right and the correspondence metamodel in between.
The source metamodel consists solely of the \house{} class and \houseType{} enum. 
Each \house{} contains information about its \houseType{} and has a reference to the next \house{} (neighbour). 
The \houseType{} determines the architecture of a \house{} on the target side, namely which construction steps are needed to build that \house{}. 
This relation will be expressed in the following using a set of graph grammar rules.
Additionally, each building contains information about its \architect{}, which has no representation in the target model.
On the target side, there are six classes. 
There is a (construction) \plan{} for every row of houses that contains the corresponding \constructions{}.
Each \construction{} contains the name of its assigned (construction) \company{} and a sequence of \constructionSteps{} consisting of \cellar{}, \floor{} and \saddleRoof{} elements that have to be processed in the given order. 
Finally, we connect both metamodels using the correspondence type depicted as a hexagon, which maps \houses{} to their corresponding \construction{}. 

\begin{figure}
	\centering
	\includegraphics[width=0.8\textwidth]{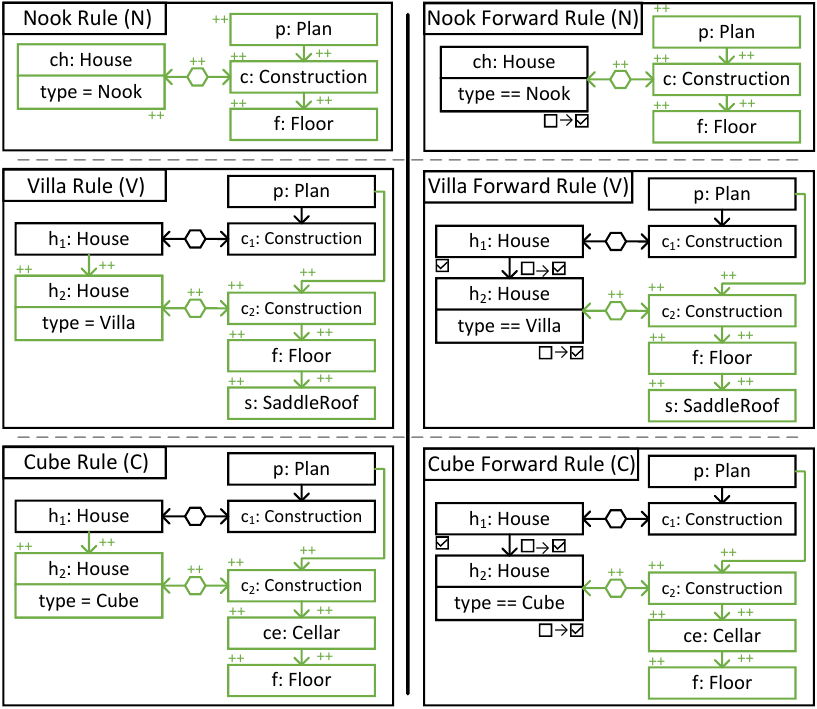}
	\caption{TGG Rules} 
	\label{fig:fundamentals:tggRules}
\end{figure}

To define a consistency relationship between source and target models, we employ \emph{Triple Graph Grammars} (TGGs)~\cite{Schurr94}, which are a declarative and rule-based way to achieve this.
As the name indicates, the rules of a TGG define the language of all consistent triple graphs, where our source, target and correspondence models are interpreted as graphs. 
\Cref{fig:fundamentals:tggRules} shows the TGG rule set of our running example, consisting of the three rules on the left.
Before a rule can be applied, its precondition must be met, meaning that all context elements in black must exist and all attribute conditions hold.
When the rule is applied, all elements depicted in green and annotated with ++ are created. %
The first rule is \nookRule{}, which can be applied without any precondition.
It creates a \nook{} \house{} together with a corresponding \construction{} on the target side and a connecting correspondence link. 
Since this will create the first \house{} of a row, it also creates a \plan{} on the target side. 
Finally, each \nook{} \house{} must have a \floor{} but no \cellar{} or \saddleRoof{}. 
The two rules \cubeRule{} and \villaRule{} are very similar to each other in that they both require a \house{} in the row that must already exist together with a corresponding \construction{} and \plan{}. 
Given that, they create a \house{} with the \cube{} or \villa{} \houseType{} as the next \house{} in line and a corresponding \construction{}, where a \cube{} has a \cellar{} and \floor{}, while a \villa{} will have a \floor{} and \saddleRoof{}.
For better readability, we do not display edge types in the rules; they can be unambiguously inferred from the metamodels. 

Using these rules, we can create consistent models from scratch. 
More interestingly, we can transform these rules to obtain forward translation rules as is depicted on the right in \cref{fig:fundamentals:tggRules}. 
The main difference to our original rules is that their created source elements have now become part of the precondition.
Assuming that we want to translate a source to a target model, this makes sense as a source model must already exist beforehand.
To avoid translating elements more than once, we introduce the annotations \marking{} and \marked{}.\footnote{These markings are not needed for the original TGG rules.} 
Source elements that were created in the original TGG rule are annotated with \marking{} in the derived forward rule.
This indicates that they mark elements upon translation such that the rule is only applicable to elements that have not been marked yet.
Other source elements that were already black are annotated with \marked{}, which means that these elements must have already been marked; different formalisations of such a marking mechanism are available~\cite{HermannEGO14,LeblebiciAFVS17,Kosiol22}.
Also, attribute assignments now turn to attribute constraints.
In our example this means that instead of setting the \houseType{} each forward rule can only be used to translate a specific \houseType{}.
\emph{Backward rules} that allow translating construction plans to the corresponding row of houses can be derived completely analogously.

\subsection{Precedence Graphs \& Synchronisation}
\begin{figure}
	\centering
	\includegraphics[width=1\textwidth]{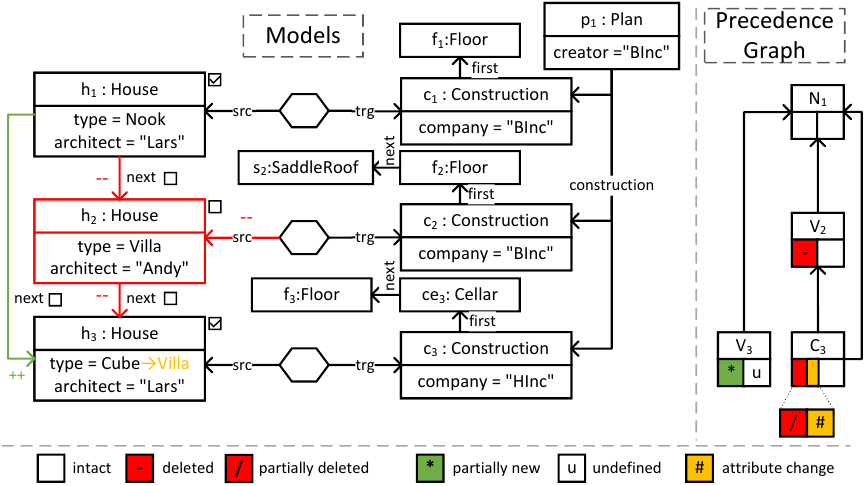}
	\caption{Synchronisation Example} 
	\label{fig:fundamentals:syncExample}
\end{figure}

We want to incrementally synchronise changes from one model to another. 
\Cref{fig:fundamentals:syncExample} depicts a small model, which was consistent w.r.t. our TGG but then was changed on the source side. 
On the left, we see a \nook{} \house{} h${_1}$, followed by a \villa{} h${_2}$ and a \cube{} h${_3}$, while on the right side there are the expected corresponding elements in the form of a \plan{} p$_1$ and three \constructions{} c${_1}$, c${_2}$ and c${_3}$, each with its \constructionSteps{}.
As a change, we deleted the second \house{} h${_2}$, which means that the third \house{} h${_3}$ now succeeds h${_1}$.
Furthermore, we changed the type of h${_3}$ from \cube{} to \villa{}.

On the right side of \cref{fig:fundamentals:syncExample}, we can see a so-called \textit{precedence graph}, which is of particular interest for the rest of this paper. 
Some of its nodes correspond to the rule applications that created the original triple graph; others represent rule applications that could have possibly been used to create the newly added elements on the source side. 
Edges denote sequential dependence between rule applications. 
Each \textit{precedence node} contains the initials of the corresponding TGG rule together with a small index, which is the same as the one of the created elements of this rule application.
Apart from that, each node has two boxes where the left and right depict the consistency state of the source and the target side, respectively.
Blank boxes indicate that the elements on this side are still intact w.r.t. the corresponding TGG rule, which means that they have not been tampered with or that the changes had no consistency violating effect, e.g., by changing the \architect{}.
Green boxes containing a \enquote*{+} indicate that new elements have been detected, which can be translated by a TGG rule, while \enquote*{*} would mean that there are new elements but they cannot be translated because some elements that would need to be marked together with them are already marked, i.e., they are still part of another (possibly inconsistent) rule application. 
In both cases, the right box for the target side is annotated with \enquote*{u}, expressing our lack of information on whether the corresponding target elements already exist.
In contrast, red boxes containing a \enquote*{-} indicate that all translatable elements on this side were deleted, meaning that created elements on the opposite side should be removed as well.
A red box with \enquote*{/} states that some but not all translatable elements were deleted, which means that the remaining elements must be translated differently than before. 
Finally, an orange box with \enquote*{\#} means that an attribute was changed such that a rule application has become invalid, e.g., the \houseType{}. 
A formalisation of precedence graphs via (partial) comatchs can be found in~\cite{Kosiol22}. 

Regarding our example, we have an intact \nookRule{} application N$_1$ on top.
The \villaRule{} application V$_2$ depends on N$_1$ as this provides the previous \house{}, \construction{} and  \plan{}.
V$_2$ itself is no longer intact as indicated by \enquote*{-} on the source side due to the deletion of \house{} h$_2$.
\cubeRule{} application C$_3$, which depends on V$_2$ is also no longer intact due to an attribute change within \house{} h$_3$ from \houseType{} \cube{} to \villa{}, which is denoted as \enquote*{\#}. 
C$_3$ is also inconsistent due to the deletion of the edge between \house{} h$_2$ and h$_3$, which is a partial deletion meaning that not all created source elements of this rule application were removed.
Partial deletions are denoted with \enquote*{/}. 
Due to this attribute change and the new edge between \house{} h$_1$ and h$_3$, we could retranslate h$_3$ using \villaRule{} as represented via the precedence node V$_3$.
However, since h$_3$ is still part of the former rule application C$_3$, we find an \enquote*{*} annotation on the source side of V$_3$; C$_3$ has to be revoked before it becomes possible to retranslate h$_3$. 

Those annotations and the actual precedence graph are constructed using the results of an incremental graph pattern matcher (IGPM) that tracks all rules\rq{} pre- and postconditions.
If a postcondition is violated, then the IGPM engine will notify us that the postcondition can no longer be matched and we can analyse which model change caused this.
For preconditions, we also track whether the source (or target) side are matched for forward (or backward) rules, which we refer to as source (or target) matches.
This gives us all possible translation steps even if the necessary context has not yet been created on the target side that would enable the translation to be executed. 
Note, however, that some steps may be mutually exclusive either because they would translate the same elements or because they rely on other rules to create necessary context on the opposite side.

\subsection{Preserving Information using Short-Cut Rules}
One sees that the above-described change of moving the third house and changing an attribute should only trigger a change of \constructionSteps{} while the information about the \architect{} should persist. 
To achieve this, we use so-called \shortcutRules{}~\cite{FritscheKST21}, which describe consistency-preserving operations, e.g., moving a house in the row without losing the target side information. 
Basically, a \shortcutRule{} revokes a rule application and applies another one instead, while preserving those elements that would be deleted just to be recreated.
For our example, a \shortcutRule{} could exchange the \houseType{} while adding the missing and removing the now superfluous \constructionSteps{}. 
A TGG \shortcutRule{} is created by overlapping an inversed TGG rule (the \textit{replaced} rule that revokes a former rule application) with another TGG rule (the \textit{replacing} rule).
Deleted elements from the replaced rule that are overlapped with created elements from the replacing rule are preserved and become context as they would otherwise be unnecessarily deleted and then recreated.
Consequently, deleted elements from the replaced rule that are not in the overlap must be removed and, analogously, created elements from the replacing rule that are not part of the overlap must be created.
Finally, only the attribute conditions from the replacing rule must hold after applying the \shortcutRule{} as we want to revoke the former replaced rule application.

\begin{figure}
	\centering
	\includegraphics[width=1\textwidth]{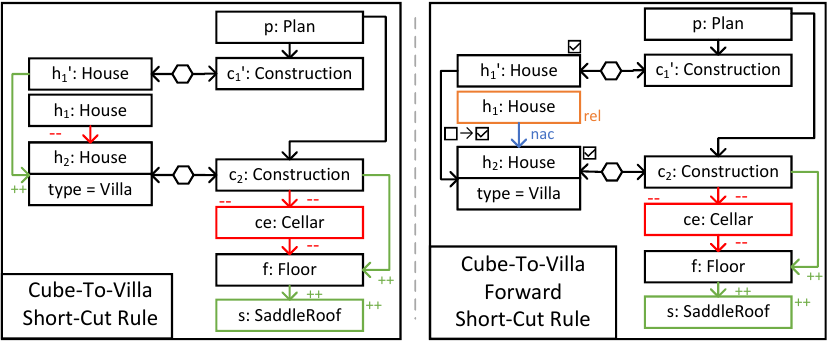}
	\caption{\cube{}-To-\villa{} \shortcutRule{}} 
	\label{fig:fundamentals:shortcutRule}
\end{figure}

The precedence graph and the dependency links between precedence nodes tell us which TGG rules to overlap with each other. 
In our example (\cref{fig:fundamentals:syncExample}), there is an invalid \villaRule{} application, while there is a new \cubeRule{} application that could be applied instead to make the now necessary transformation steps for \cube{} h$_3$.
We, thus, overlap \villaRule{} with \cubeRule{} such that the parent \house{} is not overlapped as it has changed, while the created \houses{} are assumed to be the same as well as their corresponding \constructions{}, the \plans{} and \floors{}. 
Note that generally, there are many ways to overlap rules, which means that there may be many possible \shortcutRules{}.
As discussed in \cref{chapter:contribution_extended}, we use optimisation techniques to encode the space of all overlaps and calculate a replacing rule that matches the given situation.
Note that this process is completely automated.
\Cref{fig:fundamentals:shortcutRule} depicts the resulting \shortcutRule{} on the left.
It tells us that we can move a \house{} and change its \houseType{} from \villa{} to \cube{} at the same time if, on the opposite side, we remove the superfluous \cellar{} and add a new \saddleRoof{}.

As with TGG rules, \shortcutRules{} can be transformed to yield forward and backward operationalised versions.
On the right of \Cref{fig:fundamentals:shortcutRule}, we can see its forward operationalisation.
Intuitively, we have to make sure that a user made the same changes as the \shortcutRule{} on the source side so that applying the forward operationalisation will propagate these changes correctly. 
Similar to before, formerly created elements on the source side must be marked because they are new, while black elements remain marked.
Deleted elements on the source side must have been deleted by a user change.
Hence, we must ensure that these elements do no longer exist, which is expressed as a \emph{negative application condition}\cite{EEPT06} depicted in blue and annotated with \enquote*{nac}. 
Also note that some context elements are omitted from the rule that would stem from the replaced rule, e.g., the context construction.
We can leave some of them out, if they are not needed to perform the \shortcutRule{}.
This is the  case for the \house{} h$_1$, which is needed to check whether the edge to h$_2$ has been deleted.
The \enquote*{rel} annotation on the right side indicates that h$_1$ is relaxed. 
Relaxed means that this element does not necessarily need to exist, because if it was deleted together with its adjacent edges, then the \enquote*{nac} is satisfied.
Of course, this is only reasonable if h$_1$ is indeed the \house{} that was used as context in the replacing rule application that we are going to replace.
We ensure this by only applying \shortcutRules{} at locations with formerly valid replaced rule applications and using the information about these rule applications to construct those parts of our \shortcutRule{} match that overlap with the replaced rule. 

Using \cube{}-To-\villa{} \textit{forward} \shortcutRule{}, we can now resolve the situation from before by first processing the deleted second \house{} and deleting its corresponding parts and then repairing the target side of the third \house{} by using this rule.
While the result looks very similar to the one of translating the whole source model from scratch, we still have the information about each \constructions{}\rq{} \architect{}, which means that this information is no longer lost during synchronisation.
Another advantage is that moving a \house{} in a long row of terrace \houses{} can be very expensive as all succeeding rule applications would have to be revoked.
In many cases, \shortcutRules{} can also help with this issue by preserving the consistency of subsequent steps and thereby boost the performance.

\section{Higher-Order Short-Cut Rules}
\label{chapter:contribution}
Formally, a short-cut rule is sequentially composed from a rule that only deletes structure and another one that only creates structure~\cite{FritscheKST18}. 
(We have generalised that kind of sequential composition to arbitrary rules in~\cite{KosiolT23}.) 
In practical applications so far, we made use of a static, finite set of short-cut rules; we composed each inverse of a rule from the given TGG with every rule from the TGG~\cite{FritscheKMST20,FritscheKST21,Fritsche22}. 
Intuitively, such a \shortcutRule{} replaces the action of the inverse rule by the one of its second input rule (while preserving common elements).   
Note that the \shortcutRule{} from \cref{chapter:fundamentals} is created in that way and that such \shortcutRules{} enable repairs in many situations. 
But certain complex model changes are not supported yet, namely situations in which the common effect of several TGG rules should be replaced by the effect of another set of TGG rules. 
For this, one needs \shortcutRules{} that are not just computed from the TGG rules (and their inverses) but from (arbitrary long) \emph{concurrent rules}~\cite{EEPT06}, i.e., sequential compositions, of the TGG rules (and their inverses). 
Yet, in contrast to before, we cannot precalculate this set as there are usually infinitely many ways to concatenate an arbitrarily large set of TGG rules and, thus, infinitely many \shortcutRules{}. 
Hence, we must investigate each inconsistency and deduce what TGG rules to concatenate at runtime to create a helpful \textit{higher-order} \shortcutRule{} that repairs multiple rule applications at once. 
In this chapter, we will explain the process of deriving these new rules using our running example and conclude with a discussion on its correctness.

\begin{figure}
	\centering
	\includegraphics[width=1\textwidth]{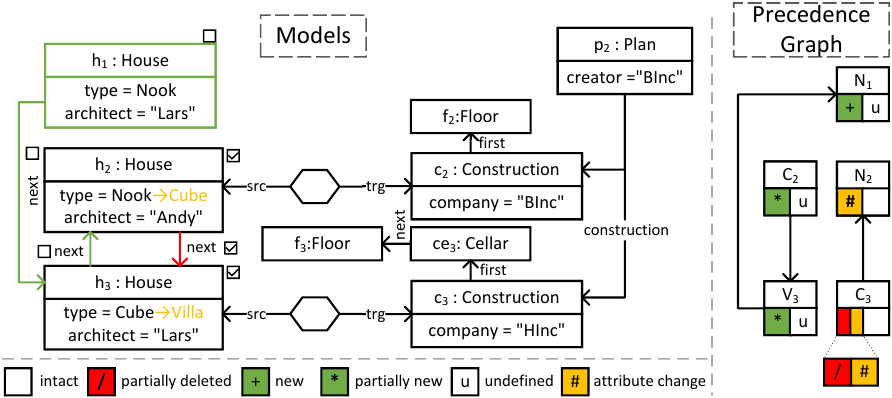}
	\caption{Synchronisation Example} 
	\label{fig:contribution:example1}
\end{figure}
\subsection{Exemplifying the Need for Higher-Order Short-Cut Rules}
\label{subchapt:motivation}
\Cref{fig:contribution:example1} shows another example model together with its precedence graph. 
We can see two \houses{}, where one was of \houseType{} \nook{} and the other of \houseType{} \cube{} together with their corresponding \constructions{}. 
Then, the first \house{}'s \houseType{} was changed from \nook{} to \cube{} and another \house{} of \houseType{} \nook{} was added at the beginning of this row. 
Similar to \cref{fig:fundamentals:syncExample}, this results in an inconsistency shown in the precedence graph on the right.
There, we see that the \nookRule{} application $N_2$ has become invalid due to an attribute constraint violation. 
Intuitively, we would expect that a new \construction{} is created for the new \nook{} \house{}, which is then added to the already existing \plan{}.
However, our consistency specification makes this rather hard to achieve as the \nookRule{} creates this \plan{} together with the first \house{} in a row.
By handling one rule application at a time, we have to translate the new \nook{} \house{}, create a corresponding \construction{} \emph{and} another \plan{}.
Then, we would either have to revoke the former \nookRule{} application and retranslate the \cube{} \house{} or use a \shortcutRule{} to transform a \nookRule{} to a \cubeRule{} application.
Since \cubeRule{} does not create a \plan{} but requires one, either way we would have to delete the old \plan{} and then try to connect the newly created or preserved \constructions{} with the new \plan{}.
While this restores consistency, this procedure has two disadvantages. 
First, any information stored in the original \plan{} would be lost and, second, we would have to fix all succeeding rule applications to connect them with the new \plan{}.
Hence, we want to create a \shortcutRule{} that preserves the original \plan{} by translating the new \nook{} \house{} and repairing the \cube{} \house{} in one step. 
Since also the order of the two original \houses{} was changed, this repair step necessarily needs to involve reacting to that and performing the required repair for the former \cube{} \house{} that now is a \villa{}. 
Hence, a higher-order \shortcutRule{} is needed. 
\begin{figure}
	\centering
	\includegraphics[width=1\textwidth]{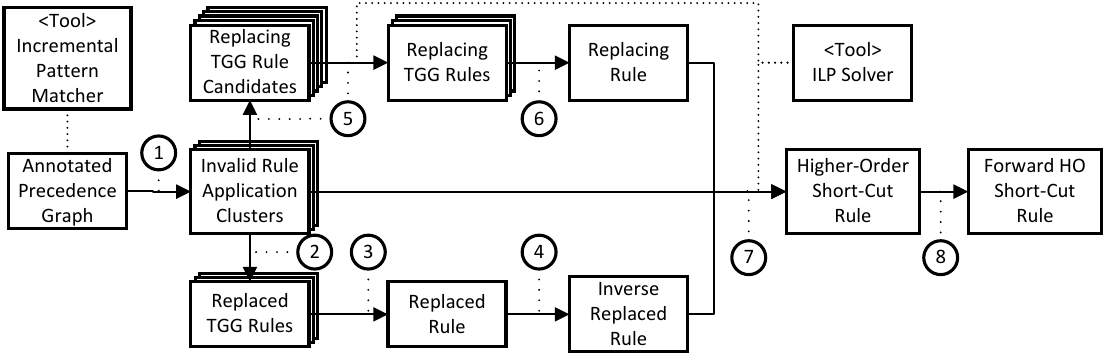}
	\caption{Construction Process of Higher-Order Short-Cut Rules} 
	\label{fig:contribution:process}
\end{figure}

\subsection{Deriving Higher-Order Short-Cut Rules On-the-fly}
\Cref{fig:contribution:process} shows our proposed process on how to obtain the necessary \textit{higher-order} \shortcutRule{} using the information from an annotated precedence graph.
Remember, to create a \shortcutRule{}, we need two ingredients: a \textit{higher-order} \textit{replaced} and \textit{higher-order} \textit{replacing} rule; i.e., one concurrent rule, built from inverses of the TGG rules, and one built from the TGG rules.
The first is trivial to obtain as we can directly derive it from neighboured inconsistencies in our precedence graph (\circled{\hyperref[fig:contribution:process]{2}}) by forming sets of inconsistent precedence graph nodes that are connected (\circled{\hyperref[fig:contribution:process]{1}}).
Such a set is found by picking an inconsistent precedence node and exploring its adjacent neighbours.
Every neighbour that is also inconsistent is added to the set and we explore its neighbours recursively. 
Note that here, and also for the synthesis of the replacing rule, the order in which we compose more than two rules is irrelevant since sequential rule composition is associative~\cite{BehrK21}. %\todo{JK: Satz ergänzt}
In contrast, nodes that are still intact or depict alternative rule applications are ignored and not explored further.
We, thereby, identify all inconsistencies that should be repaired in one step.

\begin{figure}
	\centering
	\includegraphics[width=1\textwidth]{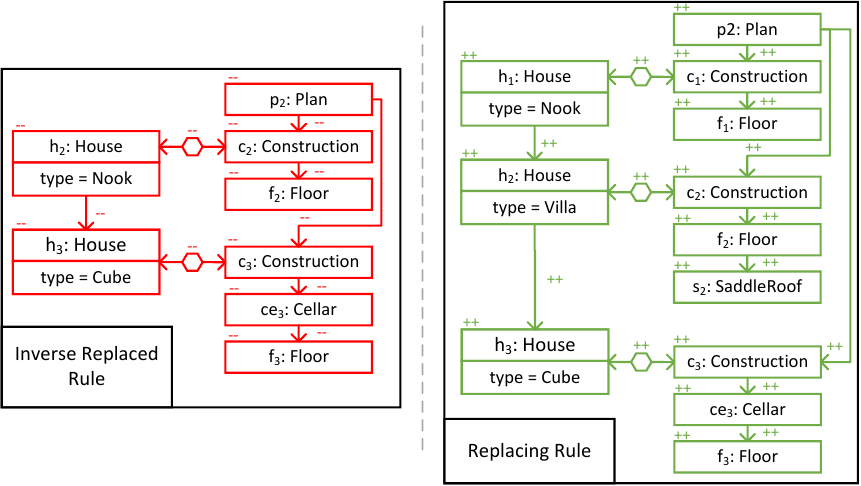}
	\caption{Inverse \textit{Higher-Order Replaced} Rule \& \textit{Higher-Order Replacing} Rule} 
	\label{fig:contribution:concatRules}
\end{figure}

In our case, we only have one set consisting of the precedence graph nodes N$_2$ and C$_3$.
These nodes represent formerly valid TGG rule applications for which we know what elements were created and which were used as precondition.
We also know what changes invalidated them.
Thus, we can use this information to concatenate \nookRule{} and \cubeRule{} (\circled{\hyperref[fig:contribution:process]{3}}) because we know that the \plan{}, \house{} and \construction{} required by C$_3$ were created by N$_2$.
In general, the resulting \textit{replaced} rule can have context elements that stem from rule applications that are still considered intact.
The resulting inverse \textit{replaced} rule (\circled{\hyperref[fig:contribution:process]{4}}) is depicted on the left of \cref{fig:contribution:concatRules} and can delete all elements created by the corresponding rules. 

The much more challenging task is to synthesise the \textit{replacing} rule. 
While we used some inconsistencies to deduce our inverse \textit{replaced} rule, we now need information that tells us how to restore consistency (\circled{\hyperref[fig:contribution:process]{5}}).
We, thus, need to use the precedence graph nodes with \enquote*{+} annotations, which indicate new translation options using entirely new elements, and those with \enquote*{*} that describe translation alternatives for elements that are still part of an older rule application.
Yet, not all of these nodes are relevant as some translations could just be carried out without using \shortcutRules{} or they might be needed for fixing another inconsistency.
We can identify the relevant nodes as those that (partially) overlap with nodes from our set of connected inconsistencies.
Intuitively, these overlapped nodes are the ones we would like to reuse and, thus, make part of a new rule application that is again consistent.
Naturally, \enquote*{+} annotated nodes cannot overlap with any inconsistencies as their source elements are entirely new.
However, at times it is necessary to repair elements and, simultaneously, translate new elements as we have seen in~\cref{subchapt:motivation}.
We, thus, also transitively add precedence nodes annotated with \enquote*{+} if they are connected with a \enquote*{*}  or \enquote*{+} annotated node that already belongs to the set of relevant nodes. 

In general, we try to identify clusters of precedence nodes that are connected and only consist of nodes with a \enquote*{*} or \enquote*{+} annotation.
The intuition behind this step is that all these nodes represent rule applications that are related and might have to be repaired in one step to yield a better result, where each cluster is used to create a specific replacing rule.
Having a set of possible rule applications, we now have to construct our \textit{replacing} rule (\circled{\hyperref[fig:contribution:process]{6}}).
However, we only have a partial knowledge on how to concatenate the identified rules because these changes have not yet been propagated.
Regarding our example, we know that we have to concatenate \nookRule{}, \villaRule{} and \cubeRule{}, but only on the source side we know exactly how.

In general, there are more challenges that we have to overcome. 
First, we have to make sure that after the repair, all elements on the source side are again part of an intact rule application.
This is a hard task because we may find that there are multiple ways to translate a specific element, e.g., by different rules that compete with each other. 
Then, selection of an appropriate set of rule applications is required. 
In some cases, this choice can influence how much information on the target side is preserved. 
Furthermore, we must select a set of rule applications such that the precondition of every selected rule application holds -- through other rules selected for this replacing rule or through original rule applications that are still intact. 
Note that for sake of brevity, our example is rather small and does not show such a scenario.
However, we discuss this case in \cref{chapter:contribution_extended}.
Finally, we must ensure that the newly repaired rule applications will not introduce a cyclic dependency, i.e., a situation where rule applications would mutually guarantee their preconditions; in our example, for instance, a cyclic row of houses. 

To guarantee these conditions, we automated the encoding of the selection of a suitable subset of to be concatenated rules and the calculation of their overlap as an ILP problem. 
A solution to such a problem will essentially describe an applicable sequence of rule applications that do not lead to a dead-end.
We encode competing rule applications as mutually exclusive binary variables and define constraints ensuring that a rule is only chosen if (one or many in combination) other rules create the elements needed by its precondition.
Hence, we have to ensure via constraints that context elements are mapped to created elements of compatible type (the same or a sub-type) and that edges can only be mapped if their corresponding source and target nodes are also mapped accordingly.
We also make sure that all remaining source elements that are part of the inconsistency must be handled by the replacing rule application. 
This encoding is inspired by former approaches that present the finding of applicable TGG rule sequences as an ILP problem, e.g.,~\cite{Leblebici18,WeidmannA20}.
Regarding our example, there are no competing rule applications but considering the precondition of \villaRule{} and \cubeRule{}, we know that the \plan{} must be the one created by the \nookRule{} application.
The \construction{} could be taken from the \nookRule{} or in case of \cubeRule{} from \villaRule{}.
Due to our knowledge that the needed \house{} of \cubeRule{} stems from a \villaRule{} application, the constraints will make it infeasible for \cubeRule{} to take the \nookRule{}'s \construction{} as there is no correspondence node between them.
The identified rules are concatenated (\circled{\hyperref[fig:contribution:process]{6}}) and form the \textit{replacing} rule as depicted on the right of \cref{fig:contribution:concatRules}.

Using both the inverse \textit{replaced} and \textit{replacing} rule as well as the information of the actual inconsistency (\circled{\hyperref[fig:contribution:process]{7}}), we can now construct the \shortcutRule{}. 
As with the search for a \textit{replacing} rule, we know what elements to overlap on the source side. 
This means that we only have to calculate the overlap for the correspondence and target side. 
While there are still many possible overlaps, e.g., between all \constructions{} of both rules, we usually want to find the one(s) that preserve more elements. 
To support least surprise, for target elements that have a corresponding source element we prefer solutions that only identify these target elements if their corresponding source elements are also identified. 
We also encode the overlapping as an ILP problem to find an optimal solution w.r.t. preserving elements.
Currently, we do not differentiate between elements but the optimisation function can be customised to favour elements that, e.g., contain attributes, which have no representation on the other side and should be prioritised. 
In our computation of overlaps, we also support node type inheritance. 

\begin{figure}
	\centering
	\includegraphics[width=1\textwidth]{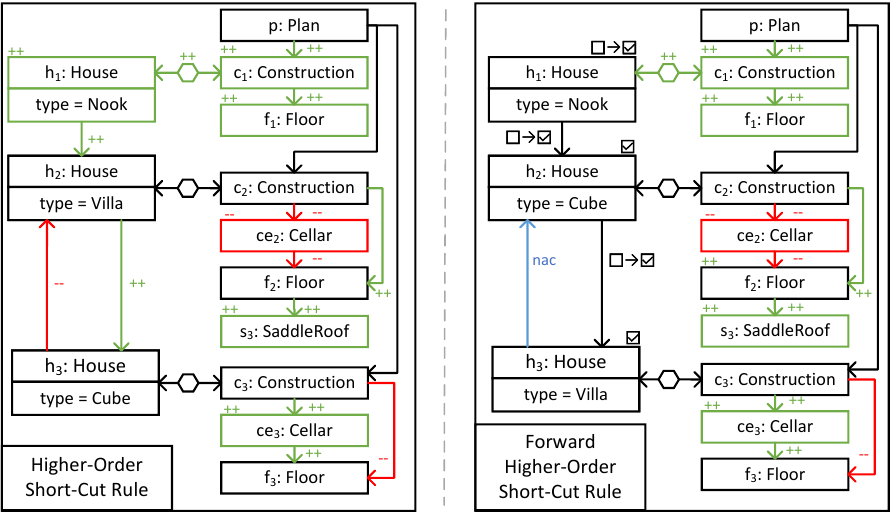}
	\caption{Higher-Order Short-Cut Rule and Forward Higher-Order Short-Cut Rule} 
	\label{fig:contribution:hoscRule}
\end{figure}

The resulting higher-order \shortcutRule{} is depicted in \cref{fig:contribution:hoscRule} on the left. 
It results from overlapping the \houses{} h$_2$ and h$_3$ from the replaced rule with h$_3$ and h$_2$ from the replacing rule.
We also overlap p, c$_2$, f$_2$, c$_3$ and f$_3$ from the replaced rule's target side with p, c$_3$, f$_3$, c$_2$ and f$_2$ from the replacing rule.
Finally, we also overlap correspondence links and edges if their corresponding source and target nodes are also mapped onto another.
By applying this rule, we can preserve the two \houses{} on the left together with their corresponding \constructions{} and \floors{}.
We also preserve the \plan{} and create a new \nook{} \house{} with corresponding target structures.
Since h$_2$ was a \nook{} \house{}, we now assign the new \houseType{} \cube{}, which means that we have to add a new \cellar{} on the target side and make it the first \constructionStep{}. 
Similarly, h$_3$ was a \cube{} \house{} but is now transformed to a \villa{} \house{}, which means that the former \cellar{} is now superfluous and thus removed.
Instead, we add a new \saddleRoof{} as a final \constructionStep{}.
As before, we can operationalise this rule  (\circled{\hyperref[fig:contribution:process]{8}}) to obtain a forward (and likewise backward) variant as depicted in \cref{fig:contribution:hoscRule} on the right.
Applying it to our example from \cref{fig:contribution:example1}, we can now translate the new \nook{}, while simultaneously repairing the other two \houses{} and preserving the \plan{} with its content.

\subsection{Discussion}
From a technical point of view, there are some aspects worth mentioning.
As a result of the construction process, we do not only get a higher-order \shortcutRule{} but also (implicitly) its match w.r.t. the analysed inconsistency.
Hence, we can directly apply the rule and repair the inconsistencies it addresses.
Besides that, however, we do not store the generated rule for applying it at other locations, which has two reasons.
First, a higher-order \shortcutRule{} is quite specific to a certain inconsistency that is characterised by a set of related broken rule applications and how exactly each of them has been invalidated.
To efficiently determine whether an existing rule can be applied to another set of inconsistencies can, thus, be very challenging and iterating and trying out all stored rules will probably become more expensive than simply constructing the rule from scratch.
Second, as a higher-order \shortcutRule{} targets multiple inconsistencies at once, the space of related inconsistencies is generally very large.
Hence, we argue that encountering the exact same constellation of inconsistencies is less likely. 
Yet, this has to be investigated further for real-world scenarios.

Regarding our choice to use ILP solving,
by applying a \shortcutRule{}, we want to preserve as much information as possible that would otherwise be lost when unnecessarily deleting and recreating elements.
Hence, we face an optimisation problem where we want to overlap as many created elements from both the replaced and replacing rule as these elements end up being preserved. 
For our current implementation, we use ILP solving separately for \circled{5} and \circled{7}. 
In \circled{5}, we want to maximise the number of mappings between context and created elements, which are the locations where the rules are glued together.
Selecting more such mappings means that less created elements must stem from outside of the computed \shortcutRule{} and, thus, it is more likely to be actually applicable.
In \circled{7}, we want to maximise the number of mappings between created elements in both rules, reducing the number of elements that must be deleted, potentially reducing information loss.
In the future, we intend to combine both problems \circled{5} and \circled{7} into a single optimisation problem because the choice of how to construct the replaced and replacing rule may affect the resulting mapping and, thus, the final \shortcutRule{}.

In the following, we shortly discuss applicability and correctness of the individual process steps. 
In \circled{\hyperref[fig:contribution:process]{1}}, we identify clusters of related inconsistencies by analysing the precedence graph. 
This information is then used to create the \textit{replaced} and \textit{replacing} rule.
Choosing the cluster too small, e.g., to save performance, might lead to a \shortcutRule{} that is not applicable if necessary context has been altered and this change was not considered.
In that case, we can still fall back to restoring consistency by revoking the rule applications and applying alternative translation steps instead, although, this has the risk of losing some information.
In contrast, if the set is chosen bigger than necessary, then the resulting ILP problems also tend to become harder to solve as the search space increases. 
In our implementation, these clusters contain all related inconsistencies that are (transitively) connected with each other without any consistent steps in between.
We, thereby, ensure that the cluster is not chosen too small as inconsistencies that are (remotely) related are repaired together.
However, at the moment, we cannot guarantee that our chosen clusters are minimal because some inconsistencies may be related but do not need to be handled together, i.e., could be repaired by standard \shortcutRules{} or smaller \textit{higher-order} \shortcutRules{}. 
In the future, we would like to investigate ways to identify situations where these clusters could be broken up to yield less complex ones, which has the potential of improving the performance of our approach. 
All \circled{\hyperref[fig:contribution:process]{2}}, \circled{\hyperref[fig:contribution:process]{3}} and \circled{\hyperref[fig:contribution:process]{4}} are trivial as we use information about formerly valid rule applications for which we rely on a pattern matcher to identify these locations.

Regarding \circled{\hyperref[fig:contribution:process]{5}}, we may encounter multiple alternative translation options for one element, which may lead to an exponentially increasing number of possible translation sequences and, thus, performance issues. 
However, under certain technical (and not too strict) circumstances, any choice can be part of a synchronisation process that finally restores consistency~\cite[Theorem~6.17]{Kosiol22} (also compare~\cite[Theorem~4]{Leblebici18}). 
Yet, as discussed for \circled{\hyperref[fig:contribution:process]{1}}, different choices may lead to different amounts of preserved model elements. 
Creating the inverse of the resulting rule in \circled{\hyperref[fig:contribution:process]{6}} is again trivial.
Regarding the underlying theory of (\emph{higher-order}) \shortcutRules{}, we refer to the literature for details~\cite{FritscheKST18,Kosiol22,KosiolT23}. 
Importantly, for our purposes in \circled{\hyperref[fig:contribution:process]{7}} and \circled{\hyperref[fig:contribution:process]{8}}, both \textit{replaced} and \textit{replacing rule} are composed from the rules of the given TGG, through our computation of \shortcutRules{} and the information provided by the IGPM, we ensure consistent matching of the replaced and the replacing rule, and we prevent the introduction of cyclic dependencies. 
This means that we meet criteria that guarantee language-preserving applications of \shortcutRules{} (cf.~\cite[Theorem~4.16]{Kosiol22} or \cite[Theorem~17]{KosiolT23}). 
The according application of their operationalised versions during synchronisation then incrementally improves consistency (compare \cite[Proposition~6.15]{Kosiol22}).

\section{ILP-based Construction}
\label{chapter:contribution_extended}
This chapter extends our previous work~\cite{FritscheKMS23}, where we could only convey the abstract concept of how to obtain higher-order \shortcutRules{}.
Our new contribution in this regard is a description on how to implement the concept using optimisation techniques and how to constrain the solution space to get desirable results.

As discussed in the previous chapter, we employ ILP solving for two subsequent tasks. 
First, we use it to find out how to concatenate a set of TGG rules to obtain the replacing rule.
Second, we use it to overlap both the replaced and the replacing rule to obtain the final \shortcutRule{}.
Note that the latter case was already presented in~\cite{Fritsche22}, which is why we will focus on creating the replacing rule. 
For that purpose, we slightly extend our example from before so that creating the replacing rule is no longer straightforward and different solutions are possible from which the most promising one should be chosen.
Also note that the entire ILP problem for this example can be found in \cref{chapter:appendix}.

\begin{figure}
	\centering
	\includegraphics[width=1\textwidth]{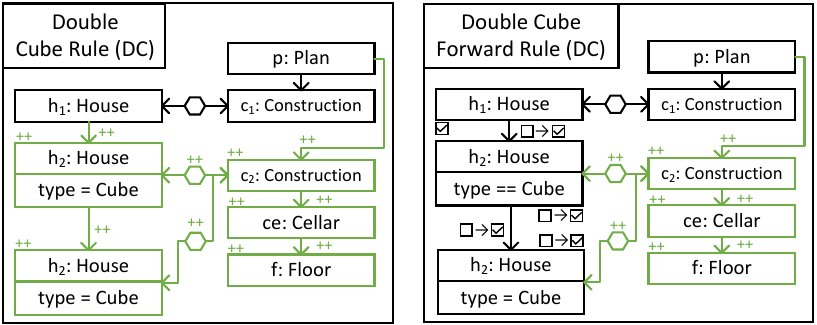}
	\caption{Double Cube Rule} 
	\label{fig:ext_Contribution:additionalRule}
\end{figure}
\Cref{fig:ext_Contribution:additionalRule} shows an additional TGG rule and its forward operationalised form that extend our rule sets from \cref{fig:fundamentals:tggRules}.
In contrast to \cubeRule{}, the new \doubleCubeRule{} creates two \cubes{} on the source side that both correspond to one newly created \construction{} on the target side meaning that both \cubes{} are planned as a semi-detached \house{}.
The purpose of this new rule is to introduce a kind of variability during the translation and likewise synchronisation of a terrace house model because there may be multiple ways on how to restore consistency now.
Translating two \cubes{} in a row, for example, we are now able to choose between applying \cubeFWDRule{} twice or \doubleCubeFWDRule{} once.
Ultimately, this is something a modeller would have to decide on their own for two reasons:
First, there is no information on the source side on whether there should be one or two \constructions{} for both \cubes{} in the target model.
Second, choosing either option will let us still translate all other \houses{} that come after both \cubes{}.
This means that this is no scenario where the translation may lead to a dead-end for certain translation sequences and, thus, neither of the two possibilities is wrong.
Of course, this means that for synchronisation scenarios, we also must account for this kind of variability when we create the replacing rule.
Specifically, we may now encounter situations with alternative TGG rule concatenations forming different replacing rules that will lead to different (but equally valid) \shortcutRules{} in the end.
For our example, this might present us with the choice between a \shortcutRule{} that uses a \doubleCubeRule{} and another using two concatenated \cubeRules{} to repair translation steps for two former \villa{} \houses{}.
Using the \shortcutRule{} based on the \doubleCubeRule{} would remove one of the \constructions{} on the other side, which could mean an unintended loss of information.
Hence, these rules mainly differ in what and how many elements are preserved and for that reason, it is important to choose the replacing rule that will be a better candidate for preserving information.

\begin{figure}
	\centering
	\includegraphics[width=1\textwidth]{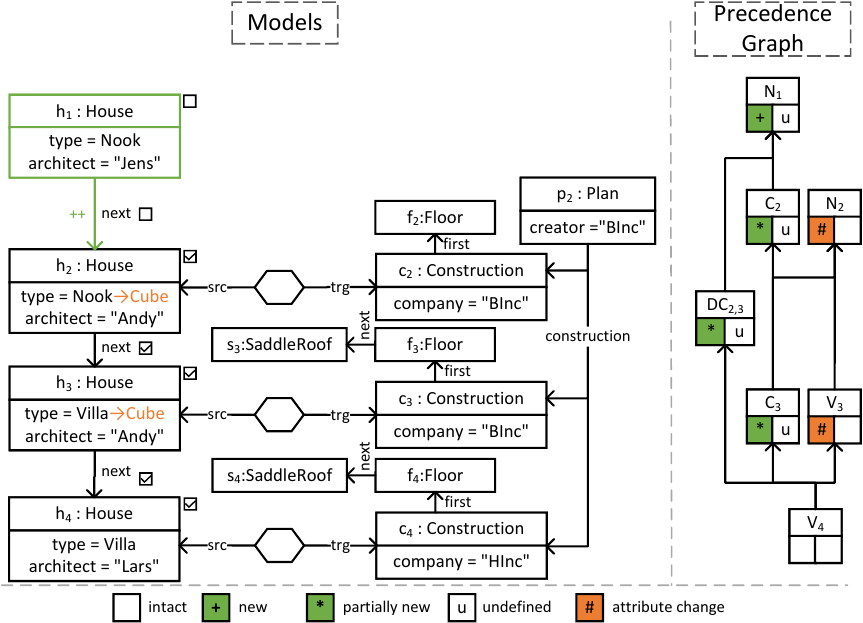}
	\caption{Synchronisation Example} 
	\label{fig:ext_Contribution:example}
\end{figure}

\Cref{fig:ext_Contribution:example} depicts a scenario in which we will encounter such a situation.
As can be seen, we start with a row of three \houses{} beginning with a \nook{} followed by two \villa{} \houses{}.
For each \house{}, there is a corresponding \construction{} as well as a \plan{} containing them on the target side.
As before, however, we change the source side and introduce an inconsistency.
This time, we change the \type{} of $h_2$ and $h_3$ from \nook{} and \villa{} to \cube{}.
Additionally, we add a new \nook{} house $h_1$ before $h_2$. 
As before, the most apparent inconsistency is that the \constructions{} $c_2$ and $c_3$ no longer contain the necessary building steps since a \cube{} has a \floor{} but no \saddleRoof{} and a \nook{} \house{} has neither.
Besides that, there are no corresponding elements for the newly added \nook{} as it has not yet been translated to the target side.
To resolve this issue and restore consistency, we must investigate the precedence graph that is shown on the right side of \cref{fig:ext_Contribution:example}.
We see that there is a new \nookRule{} application as well as an invalidated one.
The first $N_1$ is annotated with \enquote*{+} meaning that its source elements are entirely new and thus free for translation and the latter is $N_2$, which became inconsistent due to an attribute change (\enquote*{\#}). 
Since the \nook{} is now a \cube{}, we find a new \cubeRule{} application $C_2$ as an alternative translation for the \house{} $h_2$. 
Besides that, we have an inconsistent \villaRule{} application $V_3$, which was also invalidated due to an attribute change and which caused the appearance of the new \cubeRule{} application $C_3$.
In combination, changing both $h_2$ and $h_3$ to being \cube{} \houses{} lead to a \doubleCubeRule{} application $DC_{1,2}$.
Notably, the rule applications $C_2$, $C_3$ and $DC_{1,2}$ can not be applied yet, as indicated by their \enquote*{*}-annotation, because they would mark elements that are still marked by both $N_2$ and $V_3$.
Hence, our goal in the following is to create a replacing rule (followed by a \shortcutRule{}) that will let us translate the new \nook{} \house{} and synchronise the attribute change, while trying to preserve elements on the target side. 

We, therefore, have to consider several aspects.
\circled{1} We know of all elements that were part of a replaced rule application, i.e., the rule applications that were once consistent and should now be repaired.
Yet, the same does not hold for the candidates that will form the replacing rule.
This is because we only know of elements from the source domain\footnote{For a backward synchronisation case this would be the target domain.} but not the target domain, which in the precedence graph is expressed with a \enquote*{u}.
This annotation stands for undefined and indicates that we do not know yet, whether there already are any elements on this side that we could reuse or whether we should create them all from scratch.
Hence, we have to figure out whether context elements from one replacing rule candidate are created by another candidate or whether they come from yet another still consistent rule application. 
For example, we would have to figure out for $C_3$ what rule application created the \plan{}. 
\circled{2} Some candidates may exclude each other. In our case, we know that $C_2$ and $C_3$ cannot coexist with $DC_{2,3}$ because they mark the same elements on the source side.
Hence, we have to make sure that choosing a candidate means to exclude these other candidates from the solution.
\circled{3} Several \emph{dependencies} must be considered. 
There are, for example, dependencies between replacing rule candidates such as between $C_2$ and $C_3$ because the first has to be chosen (and applied) before the latter (if the latter is chosen at all).
Similar dependencies exist between rule elements, e.g., by mapping a context node to the created node of another rule because we assume that this rule creates the necessary context.
In our example, this could be the context \plan{} needed by $C_3$ and found created by $N_1$.
Edges and correspondence links are a special case in that regard as they are only allowed to be mapped when their source and target elements were also mapped onto each other, ensuring that this is indeed the same edge.
\circled{4} While there may be multiple possible replacing rules, not all candidates are equal in how much information can be preserved in the end. 
Thus, we have to find a metric for assessing the value of a solution that indicates the usefulness of a specific replacing rule w.r.t. the final \shortcutRule{}.

\subsection{Concatenation of rule candidates}
\label{chapter:contribution_extended:concatenation}
As mentioned before, we will use ILP to solve the above described problem.
In general, when using ILP, we abstractly encode search spaces by means of optimisation variables that in combination with an objective function must adhere to certain constraints.
In our case, we are only interested in 0-1 or pseudo-boolean ILP of the form:
\begin{equation}
\begin{split}
max(w^T v) \\
A v \ge b, \\
v \in \{0, 1\}^m    
\end{split}
\end{equation}
with $A \in \mathbb{R}^{m\times n}$, $b \in \mathbb{R}^n$, $w \in \mathbb{R}^{m}$ being a vector of weights and $v$ being the vector of our boolean optimisation variables $v_1, ..., v_m$.
The variables can represent entities such as rule application candidates and whether we consider them for our replacing rule or mappings between rule elements, e.g., whether we assume that a certain context element of one rule application will be created by a certain other one.
In the following, we will write $v[\ldots]$ to reference a specific variable in our vector $v$, where the value between the brackets represents a unique index.
Maximising the function should ultimately yield a valid concatenation of our replacing rule candidates, where weights can be used to favor the selection of specific variables.
In our example, all weights will usually be set to 1.
By choosing to customise these weights, however, a user can encode restoration preferences, e.g., to give certain elements a higher priority to be preserved or even penalise others that should be avoided.
Due to the various aspects that we have to consider, we will start with a small set of optimisation variables and constraints and extend them step-by-step solving the above-described sub-problems along the way.

\paragraph{Encoding rule applications and their dependencies} Constructing the replacing rule means to determine, which rule candidates to concatenate.
Therefore, for each replacing rule candidate from our precedence graph, there is one variable in $v$ that states whether this candidate is chosen for the replacing rule or not. 
For our example, we would have the following variables: $v[N_1]$, $v[C_1]$, $v[C_2]$ and $v[DC_{2,3}]$.
Then, we encode their dependencies that we can read from the precedence graph, e.g., we can see that $C_2$ depends on $N_1$, which means that $v[C_2]$ may only be set to 1 if $v[N_1]$ is 1 as well.
This can be expressed via the constraint $v[C_2] \leq v[N_1]$. 
Consequently, we introduce the same constraint for the other candidates: $v[C_3] \leq v[C_2]$, $v[DC_{2,3}] \leq v[N_1]$.
Some rule candidates are mutually exclusive because they would mark the same elements, which must be avoided because each element can only be translated once.
Hence, we need to make sure that $C_2$ and $C_3$ cannot be chosen alongside $DC_{2,3}$ and vice versa because the latter marks the same elements as both \cubeRule{} applications.
We, therefore, add the constraints $v[C_2] + v[DC_{2,3}] \leq 1$ and $v[C_3] + v[DC_{2,3}] \leq 1$, which ensure that at most one of the terms can become 1 while the other has to be 0.

\paragraph{Encoding possible dependencies of target context}
In the next step, we need to figure out how to concatenate our replacing rule candidates.
While we already know how rule application candidates can be concatenated on the source side of our example, it is not clear where target elements come from, i.e., which rule application would create them.
Hence, we have to determine whether a context element from one rule candidate might be the created element of another candidate.
For that purpose and for each candidates context (black) element, we search for created (\enquote{++}, green) elements in candidates that the current candidate (transitively) depends on and that have the compatible type and no conflicting attribute assignments/conditions.
We will call these pairs of context and created rule elements in the following \emph{context-create mappings}.
\begin{figure}
	\centering
	\includegraphics[width=1\textwidth]{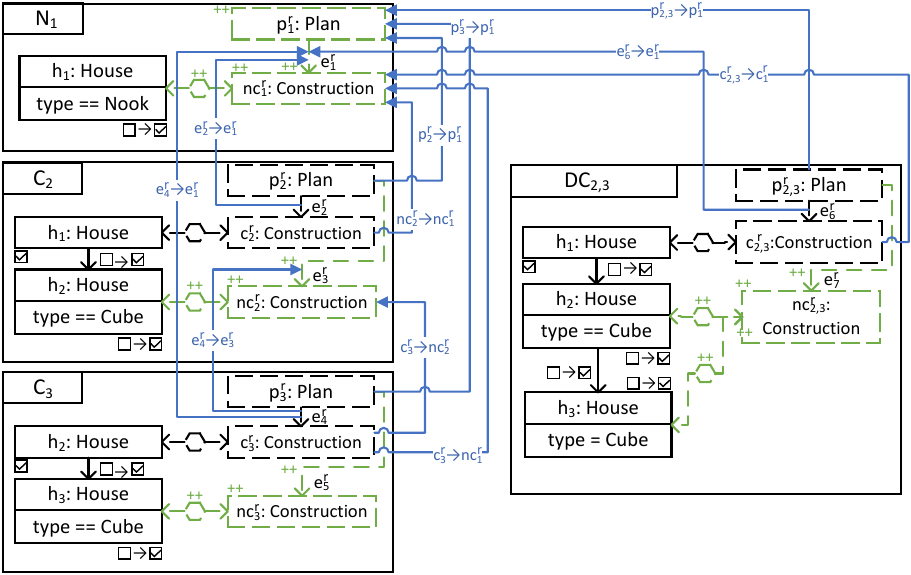}
	\caption{Context-Create Mappings} 
	\label{fig:ext_Contribution:contextCreateMapping}
\end{figure}
\Cref{fig:ext_Contribution:contextCreateMapping} depicts the context-create mappings for our rule candidates in blue, e.g., the context \construction{} $c^r_{2,3}$ of $DC_{2,3}$ could be created as $nc^r_1$ from $N_1$, which corresponds to the context-create mapping $c^r_{2,3} \xrightarrow{} nc^r_1$.
Note that the source side elements are model elements from \cref{fig:ext_Contribution:example}, while the correspondence and target side depict rule elements, which is why we draw the latter with dashed lines and added $r$ as a superscript to emphasise the difference.
In general, there may exist several possible mappings for each context element as is the case for the context \construction{} $c^r_3$ of candidate $C_3$ that could stem from $C_2$ or $N_1$.
While it is impossible that $N_1$ would create it because ultimately \construction{} $c^r_2$ will be the one that corresponds to the \cube{} \house{} $h_2$, we have no static analysis to determine this beforehand and rely on the optimiser to find a reasonable solution in compliance with all constraints.
For that reason, it is possible that no context-create mapping is chosen, which means that we assume that the context comes from outside our current scope.
In general, we cannot be sure whether our candidate set is actually able to create all necessary context or if the context stems from a still valid rule application.
Yet, incorporating all rule applications would severely impact the applicability of this approach, which would scale with the size of the model rather than the size of the change. 
That being said, we also create context-create mappings between edges.
In comparison with context-create mappings between nodes, however, edges can only be mapped if their respective source and target nodes are also mapped onto another. 
Hence, we constrain these edge context-create mapping variables to ensure that they are only 1 if the correct node mappings are 1, too.
For mapping edge $e^r_2$ from candidate $C_2$ to $e^r_1$ from $N_1$, we would need to make sure that $p^r_2$ is mapped to $p^r_1$ and likewise $c^r_2$ to $nc^r_1$.
This can be expressed as $v[e^r_2 \xrightarrow{} e^r_1] \leq v[p^r_2 \xrightarrow{} p^r_1]$ and $v[e^r_2 \xrightarrow{} e^r_1] \leq v[c^r_2 \xrightarrow{} nc^r_1]$.
Finally, we also have to create context-create mappings between correspondence links.
Note that these mappings were omitted from \cref{fig:ext_Contribution:contextCreateMapping} for readability reasons.
These context-create mappings are handled in a similar way as edges with the slight difference that we already know whether their source side matches, e.g., mapping the context correspondence link from $C_2$ to the created one from $N_1$, we can see that both are compatible on the source side because they both use the same model element $h_1$.
If this is not the case, we can directly drop the mapping and continue.

We must also add exclusions to circumvent that more than one context-create mapping is chosen for each context element.
Choosing more than one would mean that the context element is created by more than one rule application, which cannot be the case. 
As before, we can achieve this by formulating an exclusion, which ensures that the sum of all context-create mapping variables associated with the context element is less or equal than $1$, e.g., $v[c^r_3 \xrightarrow{} nc^r_1] + v[c^r_3 \xrightarrow{} nc^r_2] \leq 1$, which states that $c^r_3$ must be the element created as $nc^r_1$, $nc^r_2$ or neither.

\paragraph{Combining rule applications and context dependency}
Beyond that, we combine our rule candidate variables from before with these new context-create mapping variables to make sure that using a context-create mapping also implies that the respective rule candidates are chosen as well.
We can express this implication by stating that the variable of a context-create mapping must be less or equal than the variable of a rule candidate, which means that the first can only become 1 if the latter is 1 as well.
In our example, this must hold when setting the mapping variable $v[c^r_3 \xrightarrow{} nc^r_2]$ to 1, which means that both $v[C_3]$ and $v[C_2]$ must be 1 as well because they use elements from these candidates.
Hence, we add the constraints $v[c^r_3 \xrightarrow{} nc^r_2] \leq v[C_2]$ and $v[c^r_3 \xrightarrow{} nc^r_2] \leq v[C_3]$, where the context-create mapping can only be 1 if both $v[C_2]$ and $v[C_3]$ are 1, too. 

\paragraph{Optimisation goal}
Assuming that we would now maximise the number of chosen context-create mappings, we would get a valid replacing rule that overlaps context and created elements as much as possible.
Our constraints ensure that no competing rule candidates may be chosen at the same time, while our constraints for edge and correspondence context-create mappings ensure that a solution is favored that does not blindly map each context node to any possible creating node.
Instead, it should always give a higher reward when also considering how each node is connected.
Thereby, the edge and correspondence constraints become the glue that pushes us towards a concatenation that is not only valid but also reasonable and applicable. 

\subsection{Incorporating former target side elements}
\label{chapter:contribution_extended:modelmapping}
The result can still be improved by incorporating information about target side elements that were created by our formerly valid replaced rule applications.
When being presented with different alternatives as in our example, incorporating knowledge about formerly created elements can help us making a better decision, e.g., to preserve more elements.
For example, considering our example from above, it could be better to use a replacing rule that includes both \cubeRule{} candidates instead of the \doubleCubeRule{} because we want to repair a \nookRule{} and a \villaRule{} application that both created a \construction{}.
Choosing the \doubleCubeRule{} would result in a \shortcutRule{} that removes one \construction{}.
Even if this is what we wanted, we need the means to state that either solution should be preferred.
We can achieve this by introducing \emph{rule-model mappings} between created rule elements from our rule candidates and model elements that were created previously by rule applications that are no longer intact and which we are trying to repair with the final \shortcutRule{}.
The intuition behind this step is that we try to find out whether we can reuse and, thus, preserve elements created by prior rule applications that would otherwise be deleted.
\begin{figure}
	\centering
	\includegraphics[width=1\textwidth]{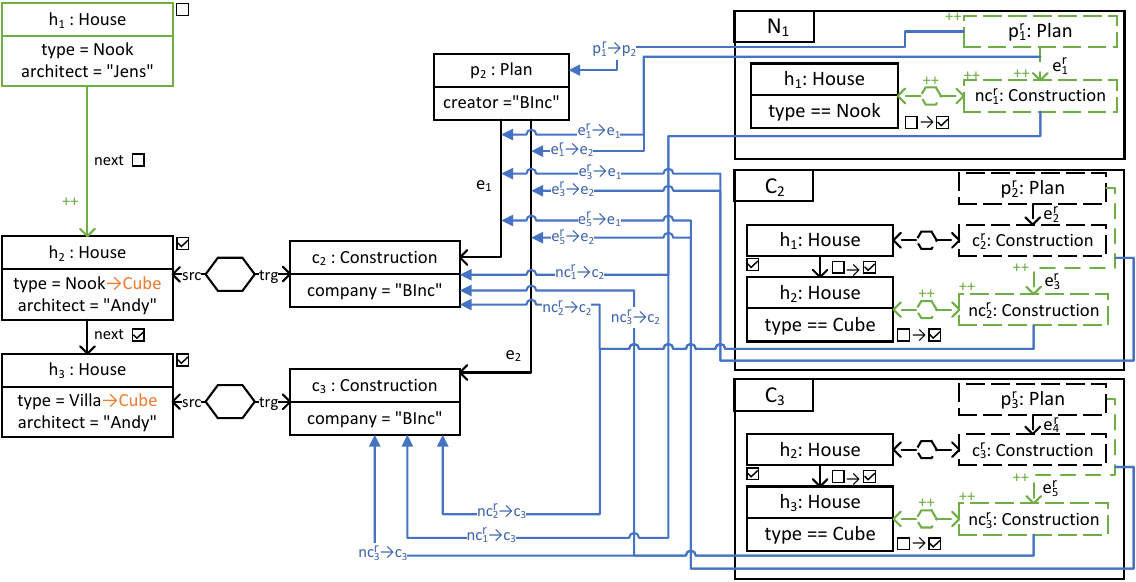}
	\caption{Mapping Rule Candidate to Model Elements} 
	\label{fig:ext_Contribution:rule2model}
\end{figure}
\Cref{fig:ext_Contribution:rule2model} depicts these rule-model mappings for our example, although, the rule candidate $D_{1,2}$ has been omitted for readability reasons.
Similar to the context-create mappings, for each context element in our rule candidates, we search for compatible model elements.
This time, however, they must have the same type (i.e., we exclude subtyping) or else the created rule element would not be the same as the one we already have. 
Again, correspondence mappings are omitted due to readability reasons but we discuss them in the following, too.
Also note that, as explained previously, we do not need mappings to source side elements, because we already know these mappings.

Having a look at \cref{fig:ext_Contribution:rule2model}, we can see that  there is only one available rule application candidate on the right side that creates a \plan{}, which means that we can only map $p^r_1$ to $p_2$.
In contrast, each of the three \constructions{} on the right side could be mapped to any of the two existing \constructions{} in our model, which results in six mappings.
This also holds for the edges between the \plan{} and a \construction{} for which we know that there must be three in the end but for which there are only two candidates in our model.
We also map created correspondence links for which there are also six possible mappings.
Creating all these mappings is an important difference between context-create and rule-model mappings.
For the first, we also considered dependencies between rule candidates and (implicitly) filtered some mappings, while for the latter, we are not sure which rule application would be a good candidate for preserving the information.
This is because the order in which elements are translated may have changed as is the case for the example in \cref{chapter:contribution}.
Hence, we create mappings from each created rule element to all matching model elements and let the optimiser find a solution that reuses these elements in a reasonable way.
Comparing the former rule applications with the new rule application candidates and incorporating this knowledge to improve performance is left to future work.

\paragraph{Encoding dependencies between rule-context mappings}
Since each model element can only be created by one rule element, we must ensure that mappings pointing to the same model element are mutually exclusive.
Similar to \cref{chapter:contribution_extended:concatenation}, we enforce this by adding a constraint that ensures that only one mapping variable can be set to one, e.g., for mappings to $c_3$ we add $v[nc^r_1 \xrightarrow{} c_3] + v[nc^r_2 \xrightarrow{} c_3] + v[nc^r_3 \xrightarrow{} c_3] \leq 1$.
Of course, these mappings may only be chosen if the rule application candidate on the right side is chosen as well.
As with the rule concatenation constraints, we can express this by stating that the rule-model mapping variable may only be 1 if the rule application candidate variable is 1, e.g., $v[nc^r_3 \xrightarrow{} c_3] \leq v[C_3]$. 

Finally, we have to make sure that edges and correspondence links are not mapped if their sources and targets are not mapped onto each other as well since edges can not be the same if they point to different elements.
In comparison to the \cref{chapter:contribution_extended:concatenation}, where we mapped context to created rule elements, we are now only interested in mapping created rule elements to formerly created model elements.
However, most of our created rule edges have one context node as source, e.g., $e^r_3$.
This means that we can only take such an edge if its context rule node is mapped to a created rule node of another rule application candidate, which is then mapped to the corresponding element in the model.
For mapping $e^r_5$ to $e_2$, we would add the following constraint: $3 v[e^r_5 \xrightarrow{} e_2] \leq v[nc^r_3 \xrightarrow{} c_3] + v[p^r_1 \xrightarrow{} p_2] + v[p^r_3 \xrightarrow{} p^r_1]$. 
It states that $v[e^r_5 \xrightarrow{} e_2]$ can only be 1 if the three other variables are set to 1 and sum up to 3, which means that $nc^r_3$ and $p^r_1$ were mapped to $c_3$ and $p_2$, respectively, while the context \plan{} from $C_3$ is assumed to be created by $N_1$.
Only then, we can be certain that $e^r_5$ could indeed create $e_2$ if  $C_3$ is chosen. 
As in \cref{chapter:contribution_extended:concatenation}, correspondence links are handled similarly as edges with the slight difference that we already know what element they point to on the source side. 
Hence, we only have to encode constraints for the target side elements of correspondence links.

Incorporating edges and correspondence links is crucial as this will ensure that nodes are not reused randomly.
Instead, those solutions will be preferred where relations between model nodes are preserved.
In our example, we could map the created \constructions{} of $C_2$ and $C_3$ freely but when mapping $nc^r_3$ to $c_2$, we won't be able to map the correspondence link and, thus, get a smaller reward.
Another interesting case is that of $N_1$.
We could not only map its \plan{} $p^r_1$ to $p_2$ in our model but also map $nc^r_1$ to $c_2$.
Doing so, however, would leave us with no option but to remove the correspondence link between $h_2$ and $c_3$ because we cannot (and probably also do not want to) repurpose the edge to point to $h_1$.
This means that edges and likewise correspondence links are our means of reasoning on whether elements are assumed to be the same and, thus, should be preserved.
Maximising the number of chosen mappings, finally, tells us how to concatenate some of the rule candidates in order to obtain our final replacing rule.
For our example and without choosing specific weights, we would choose $N_1$, $C_2$ and $C_3$ as our candidates, where the context \plan{} of the latter two stems from $N_1$ as well as the context \construction{} from $C_2$, while $C_2$ creates the \construction{} that is needed by $C_3$.
This means that the \constructions{} from $C_2$ and $C_3$ were mapped to the model \constructions{} $c_2$ and $c_3$, respectively.
\begin{figure}
	\centering
	\includegraphics[width=1\textwidth]{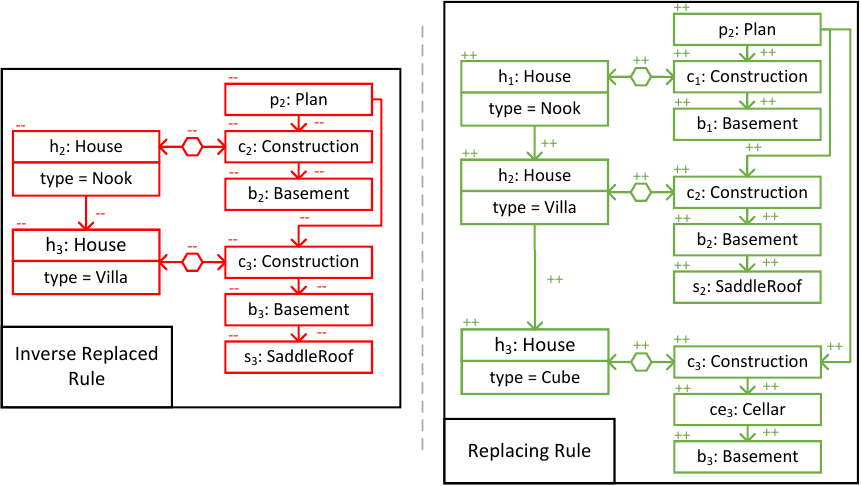}
	\caption{Replaced and Replacing Rule} 
	\label{fig:ext_Contribution:replacingAndReplacedRule}
\end{figure}
In contrast, when choosing $DC_{1,2}$, we would only be able to map one \construction{} from our rule application candidate to our model and there are less context-create mappings from context to created elements, which in total gives a smaller reward.
If we would like to prefer the $DC_{1,2}$ nonetheless, then we can choose specific weights to make it more rewarding to choose it.
An obvious choice would be to add a large weight to $v[DC_{1,2}]$ so that the reward of choosing a \doubleCubeRule{} is always higher than choosing two \cubeRule{} candidates.
As a rule of thumb, we must choose a weight that is greater than the difference between the number of mappings a \doubleCubeRule{} and two \cubeRules{}{} may have.
Another, more fine-granular, option is to add weightings to rule-model mappings such that choosing mappings between a \doubleCubeRule{} candidate and a model is worth more than choosing similar mappings for two subsequent \cubeRules{}.
The difference between the first and second option is that the latter lets us prioritise specific elements over others.

\begin{figure}
	\centering
	\includegraphics[width=1\textwidth]{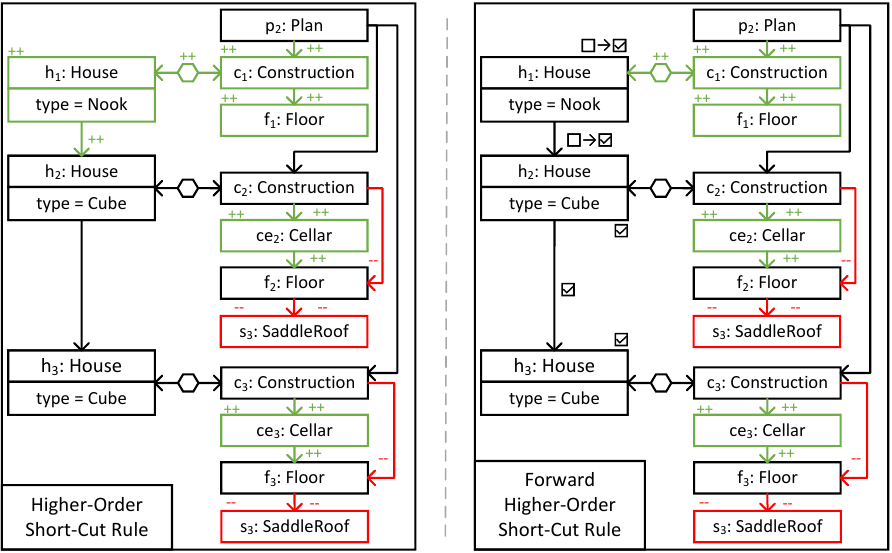}
	\caption{Higher-Order Short-Cut Rule and Forward Higher-Order Short-Cut Rule} 
	\label{fig:ext_Contribution:ho_rule}
\end{figure}

\Cref{fig:ext_Contribution:replacingAndReplacedRule} shows the resulting replaced and replacing rule.
Both are used to create the \shortcutRule{} shown in \cref{fig:ext_Contribution:ho_rule} (cf.~\cite{FritscheKST21}) by overlapping both rules and thereby identifying common elements.
Here, the overlap would be between elements with the same name.
Overlapped elements are preserved when transforming from our formerly valid replaced rule application to the replacing rule application.
Finally, applying this rule on our changed model from \cref{fig:ext_Contribution:example} would create the missing \construction{} for \house{} $h1$, while preserving the already existing \plan{} and changing the other \constructions{} to have a \cellar{} instead of a \saddleRoof{}.

\subsection{Discussion}
We would like to highlight and discuss some details of our approach.
First, we would like to point out that, in general, there are multiple possible \shortcutRules{} as there are often various possible overlaps between a replaced and a replacing rule.
While we do not search for all possible \shortcutRules{} due to performance reasons and because many of them are not useful, we limit ourselves to finding useful ones, according to a custom (hand-crafted) metric, i.e., a specific choice of weightings (cf.~\cite{FritscheKMS23}).
This means two things. 
First, in order to preserve specific elements, we need to encode this via weights such that a solution that focuses on these elements is preferred.
Second, since we do not search and try all \shortcutRules{}, our approach is an heuristic where in its basic form we choose all weights to be 1 due to our experience with examples from our repository of test scenarios\footnote{\url{www.github.com/eMoflon/emoflon-ibex-tests}}.

Of course, since our approach is a heuristic, it can happen that we are not able to find a \shortcutRule{} that can be used in that particular scenario.
As discussed in the previous chapter, however, we can guarantee that each created higher-order \shortcutRule{} is indeed correct, meaning that these rules cannot be used to perform synchronisation steps that inflict new inconsistencies.
If no rule can be found, there is still the option to retranslate elements that were part in prior rule applications, which is guaranteed to restore consistency~\cite{Leblebici18}.

With the current ILP formulation, mapping an element from a rule to a model gives some reward.
In contrast, we do not penalise if not all elements from a rule have a matching counterpart in the model, which in some cases would be reasonable.
A future extension could thus introduce such a penality term, e.g., adding another rule-model mapping variable for each rule element that has a negative weight and enforcing that for each node exactly one rule-model mapping variable has to be chosen.
If there are not enough elements in the model to map all rule elements for a rule candidate, we would have to set variables with a negative weight to 1, which reduces the overall reward.
Then, it would become more likely that another alternative candidate is chosen that explains more model elements. 

Another important aspect is that after finding the replacing rule, we create the \shortcutRule{} using ILP solving in a subsequent step with our former \shortcutRule{} framework presented in \cite{FritscheKST21}.
In fact, having found mappings from our replacing rule candidates to both the replaced rule and the model, we could already create the \shortcutRule{}.
Currently, we rely on the former construction mechanism of \shortcutRules{} but we plan to replace it with one that creates replaced, replacing and \shortcutRules{} in one step in the future.

\section{Evaluation}
\label{chapter:evaluation}
We implemented our approach and integrated it into the state-of-the-art graph transformation tool eMoflon\footnote{\url{www.emoflon.org/}}.
Our approach extends our previous work on \shortcutRules{}~\cite{FritscheKST18,FritscheKST21} and while the approach has the potential to preserve more information during synchronisation, we have to investigate how our original \shortcutRules{} perform in comparison.
Also, since our approach analyses dependencies and creates \shortcutRules{} on-demand at runtime, we have to investigate whether it introduces a noteworthy offset.
Thus, we pose the following research questions: 
\begin{itemize}
    \item \textbf{(RQ1)} Does the new runtime analysis introduce any additional cost in comparison with using a precalculated set of \shortcutRules{}?
    \item \textbf{(RQ2)} Does our new approach indeed preserve more information in certain cases?
\end{itemize}  

\paragraph{Evaluation scenarios} 
To answer our research questions, we investigate five different scenarios of our running example with three different resolution techniques:
The \textit{legacy} algorithm~\cite{Leblebici18} revokes invalid rule applications and retranslates parts of the model, while \textit{SC} stands for the original \shortcutRule{} framework with precalculcated repair rules~\cite{FritscheKST21} and \textit{HO SC} stands for our new \textit{higher-order} \shortcutRules{}. 
We compare both \textit{SC} and \textit{HO SC} with \textit{legacy} to show the impact of repairing models instead of retranslating them (partially).
Note that in both \textit{SC} and \textit{HO SC} our algorithm falls back to using the \textit{legacy} algorithm in case no \shortcutRule{} was found. 

In the first and second scenario, we have a fixed number of 2\,000 rows of \houses{} with around 26\,000 nodes on both source and target side and we increase the number of applied changes in steps of 50 from 50 to 250.
Our primary goal for these scenarios is to investigate how our approach copes with linearly scaling a number of different kinds of changes. 
In contrast, the third and fourth scenario comprise two long rows of \houses{} that we grow iteratively by adding new \houses{} in steps of 20 from 20 to 200.
This change is expected to become more expensive to resolve as the rows grow larger but should also be able to preserve more information than the other two algorithms. 
The fifth scenario was not part of our previous work and extends our former evaluation.
It features one long row of 1\,000 \houses{}, where the first \house{} is of type \nook{} and all following ones are of type \villa{}.
Then, we change an increasing number of \houses{} in this row to become of type \cube{} starting from 10 to 60 subsequent \houses{} in steps of 10.
Both the former \shortcutRule{} and the newer higher \textit{higher-order} \shortcutRule{} framework should be able to preserve the same amount of information, the latter is expected to be more expensive due to the prior analysis needed to find a suitable rule. 
We, thus, want to measure the effect that subsequently dependent changes have when using our new approach.

The specific changes for each scenario and their expected outcomes are as follows:
In the first and third scenario, we add new \nook{} \houses{} at the start of each row.
\textit{HO SC} can synthesise a \emph{higher-order} \shortcutRule{} to react to this change, while \textit{legacy} has to re-translate each affected row and \textit{SC} has to separately translate the newly created \nook{} \houses{} and then move the affected \constructions{} to the newly translated \plan{} (using precalculated \shortcutRules{}).
In the second and fourth scenario, we relocate subrows.
This is a worst-case scenario for \textit{HO SC}. 
In this case, every moved \house{}'s \construction{} must be repaired and moved to the new \plan{}.
While \textit{legacy} has to perform this via retranslation, both \textit{SC} and \textit{HO SC} can use the same repair rules -- only \textit{SC} does not have the offset of creating them on-the-fly.
In the fifth scenario, we change an increasing number of \villa{} \houses{} to being \cube{}, starting from the front.
As a result, \textit{HO SC} will have to construct larger \shortcutRules{}, which constitutes another worst-case scenario, while \textit{legacy} will simply retranslate the \houses{} and \textit{SC} can repair them step-by-step.

\paragraph{Experimental setup}
We measure the time it takes to resolve the inconsistencies as well as how many elements are deleted in the process.
Every measurement was repeated five times and we show the average runtime while the number of deleted elements did not fluctuate.
The evaluation was executed on a system with an AMD Ryzen 9 3900x with 64GB RAM.
It can be replicated using our prepared VM\footnote{\url{www.zenodo.org/record/10377155}}, which comes with a detailed explanation on how to reproduce the results.
The current version v1.1 of our VM needs a license of the Gurobi\footnote{\url{www.gurobi.com}} ILP solver to execute the evaluation but the former version is still available and implements the first four test scenarios without this dependency. 

\begin{figure}
	\centering
	\includegraphics[width=1\textwidth, trim={0cm 0.8cm 0cm 0.47cm},clip]{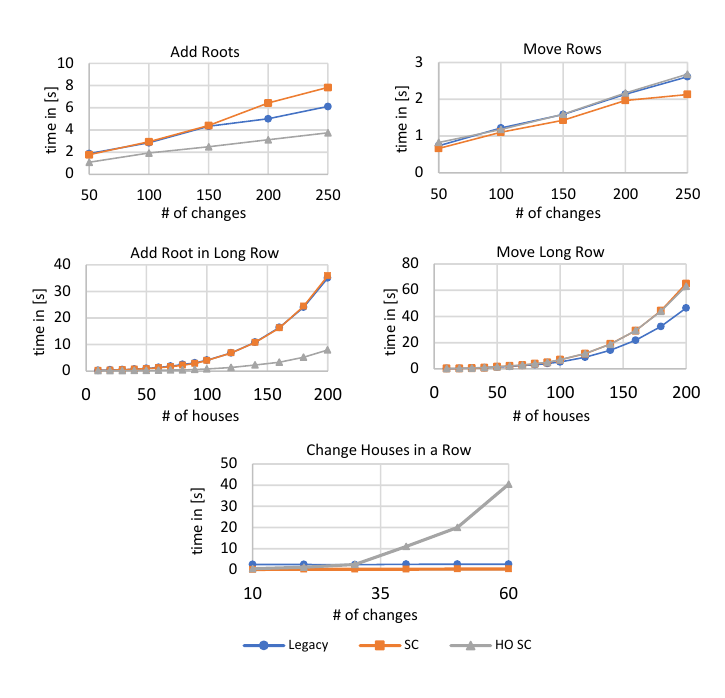}
	\caption{Time Comparison} 
	\label{fig:evaluation:timeMeasurements}
\end{figure}

\paragraph{Results}
\Cref{fig:evaluation:timeMeasurements} shows the measured times of the all scenarios.
For the first and third scenario, where \textit{legacy} and \textit{SC} have to transitively retranslate/repair, creating the needed higher-order \shortcutRules{} on-the-fly even has performance benefits. 
For the second and fourth case, the results indicate that for one of the worst-case scenarios where \textit{SC} will have the same effect as \textit{HO SC}, constructing higher-order \shortcutRules{} and applying them instead of performing many smaller repair steps only introduces a small cost.
However, the fifth scenario shows there are cases where the construction of large \shortcutRules{} can be very expensive, at least when compared to applying smaller \shortcutRules{} instead.
For changing 60 \houses{}, \textit{HO SC} takes over 40 seconds to construct a \shortcutRule{}, where over 90\% of the time is needed to find a solution for the ILP problem.

\begin{figure}
	\centering
	\includegraphics[width=0.9\textwidth, trim={0cm 0.5cm 0cm 0.6cm},clip]{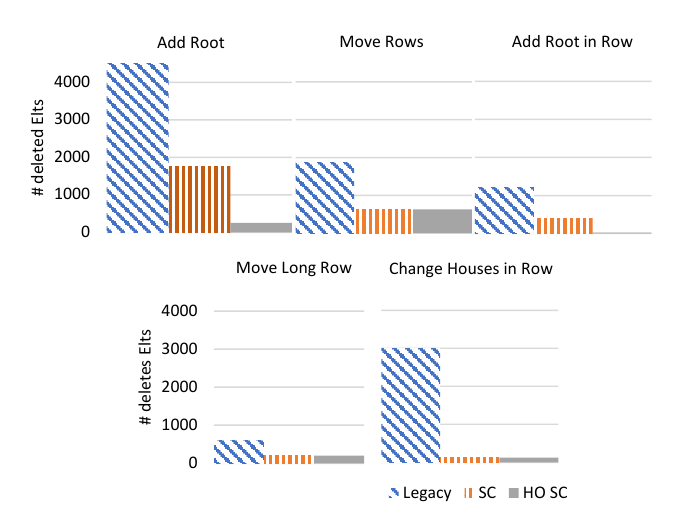}
	\caption{Comparison of deleted elements} 
	\label{fig:evaluation:deletedElements}
\end{figure}

\Cref{fig:evaluation:deletedElements} shows the largest number of deleted elements (nodes and edges) for all scenarios. 
As can be seen, using \textit{HO SC} has the potential of preserving more information than \textit{SC} and \textit{legacy} in the right cases \textbf{(RQ2)}.
In the others, both \textit{SC} and \textit{HO SC} outperform the \textit{legacy} algorithm; this cannot preserve any information at all. 

In sum, the research questions can be answered as follows:
\textbf{RQ2} can be answered with a clear yes because there is no case where our new approach preserved less information than the former \shortcutRule{} framework and for two scenarios it even preserved more.
Regarding~\textbf{(RQ1)}, we can state that there are cases where our new approach introduces a significant overhead compared to our previous \shortcutRule{} framework. 
Yet, at the same time, there are cases where using our new framework can even save time in situations where we cannot repair a model using smaller \shortcutRules{} and large parts have to be retranslated to restore consistency.
For several cases, both our new and the old \shortcutRule{} framework even performed similarly w.r.t. the performance.
Considering that using \textit{higher-order} \shortcutRules{} can indeed preserve more information, one can argue that it is advisable to use \textit{higher-order} \shortcutRules{} and switch to the former algorithm in case that a specific scenario only consists of worst-case scenarios such as the fifth scenario of our evaluation. 

\paragraph{Discussion}
There are several points worth mentioning. 
First, the current evaluation features only one example with synthetic data.
Our approach, thus, has to be evaluated further on other (preferably real-world) scenarios with non-synthetic data and user changes.
Yet, based on our experience with our test suite of 22 TGGs\footnote{\url{https://github.com/eMoflon/emoflon-ibex-tests}}, we argue that the investigated scenarios are representative and similar results are to be expected there as well.

Second, while using \textit{HO SC} has the potential to preserve more information than the other two approaches, there are cases where calculating a \shortcutRule{} would take too much time.
This issue could be resolved by not creating one large \shortcutRule{} to handle all changes at once but rather splitting them up.
Therefore, we need to find some criteria to judge whether a split could cause information loss.
Apart from that, we could also reduce the size of the ILP problem by filtering those mapping variables that cannot be part of the final solution.
Currently, we encode most mappings and rely on our ILP constraints to filter these ones out.

\section{Related Work}
\label{chapter:relatedWork}
In this section, we relate our approach to other sequential model synchronisation approaches.
We will first focus on TGG-based approaches and then discuss other related works with a different methodology.

\paragraph{TGG-based approaches}
There have been several works that implement model synchronisation by transitively revoking invalid rule applications and retranslating possibly large parts of a model~\cite{GieseW09, LauderAVS12, LeblebiciAFVS17}.
While these approaches come with proofs of termination, correctness and completeness, they may lead to an unnecessary information loss and at times a decrease in performance.

In contrast, Hermann et al.~\cite{Hermann2015} tried to solve this issue by calculating the maximal still-consistent sub-model, which is used as a starting point to propagate the model changes.
Their approach is shown to be correct but cannot guarantee that all changes are carried out in a least-changing way.
Above that, calculating the sub-model is computationally very expensive and their approach is limited to a deterministic TGG rule set only.

A very similar approach to ours is given by Giese and Hildebrandt~\cite{Giese2012} where they propose repairing rules that closely resemble \shortcutRules{}.
These also save nodes instead of unnecessarily deleting them and are able to propagate relocations of model elements.
However, neither a construction scheme to obtain these rules is provided nor is the preservation of information proven to be correct.
Yet, using our approach from~\cite{FritscheKST21}, we are indeed able to construct the same rules as mentioned in their work.
Similarly, Blouin et al.~\cite{BlouinPDSD14} also propose to use custom repair rules.

Another approach was initially proposed by Greenyer et al.~\cite{Greenyer2011} and later formalised by Orejas and Pino~\cite{Orejas2014}.
They propose that elements are never directly deleted but rather marked.
Thereby, these elements are still free to be re-used during synchronisation, e.g., by letting a forward rule transformation take elements from the set of these elements instead of creating new ones and, thus, prevent them from being deleted.
Yet, to the best of our knowledge, their approach was never implemented. 

Finally, Anjorin et al.~\cite{Anjorin2015a} discussed design guidelines for TGGs that lead to less transitive dependencies between rule applications and can, thus, be synchronised without triggering a large cascade of retranslation steps.
Yet in many cases, remodelling a TGG to comply with these guidelines means to change the defined language and may for that reason not be desired.

In summary, there are several TGG-based approaches that acknowledge the limitations of propagating changes by retranslating possibly large parts of a model.
Some approaches follow a similar approach as ours, yet without a formal foundation or construction concept.
Others have, to the best of our knowledge, never been implemented.
While we provide both, we extend our previous works~\cite{FritscheKST18, FritscheKST21} by broadening the practical applicability and preserving information in more complex scenarios than before.

\paragraph{Bidirectional transformations}
Another popular formal framework for bidirectional transformations are lenses.
While there are many works related to lenses and least-changing incremental synchronisation~\cite{Diskin2010, Hinkel2019, Hofmann2012, Horn2018, Wang2011}, most are only theoretical and were not implemented.
Most closely related to our work are the works of Hofmann et al.~\cite{Hofmann2012} and Wang et al.~\cite{Wang2011}.
Hofmann et al. introduced so-called \emph{edit lenses} that focus on the changes and how to propagate and preserve them instead of taking a state-based approach.
This is similar to our approach that works incrementally based on the perceived change to the model.
Wang et al. too derive functions to propagate changes between models.
However, their approach is limited to tree-like structures only.

Another approach by Macedo and Cunha~\cite{Macedo2016} was influenced by the lenses framework and is based on ATL and SAT solving via Alloy.
In their work, they describe how to encode the search space of all feasible synchronisation solutions as an optimisation problem. 
Their approach rates changes based on one of two metrics, namely the \emph{graph edit distance} and the \emph{operation-based distance}.
Using these ratings, a least-changing propagation is calculated, interpreting least-change either as the minimal set of (propagation) operations needed or the minimal set of atomic graph edit steps to be taken.
In comparison to our work, they incorporate more advanced metamodel constraints, e.g., OCL, and, generally, their approach should yield good results.
The major drawback is the scalability as it is applicable to rather small models only.

Finally, Anjorin et al.\cite{AnjorinDJKLW17} compared three existing state-of-the art tools for bidirectional transformations of different methodologies against each other. 
Specifically, they compared the rule- and TGG-based tool eMoflon\cite{Weidmann2019}, the constraint-based tool mediniQVT~\footnote{mediniQVT is not maintined and no longer available.}
and BiGUL~\cite{Ko2016}, which is a bidirectional programming language and related to lenses.
Comparing these tools, they showed that eMoflon tends to perform better than the other two approaches w.r.t. performance.
Additionally, eMoflon only failed on one scenario, while mediniQVT failed in four and BiGUL in two.
Using our previous smaller \shortcutRules{}~\cite{FritscheKST18}, we were already able to solve the last case and as shown in our evaluation, the new approach is able to preserve information in even more cases.
This benchmark was later extended~\cite{AnjorinBWDKEHSZ20}, featuring more scenarios and being implemented by a more diverse set of tools.
There, eMoflons performance was showing a worse runtime performance than most other approaches and was not able to solve 6 out of 34 cases.
However, an older version was used that did not incorporate \shortcutRules{} and worked with another incremental pattern matcher that has since been replaced by a more efficient one.

\paragraph{Model repair}
Model repair~\cite{MacedoTC17} is the task of restoring the consistency of a model with regard to some specification. 
Since modelling languages like UML~\cite{OMG2017} can be used to describe a single system using a set of models of different types, e.g., class and sequence diagrams, there are several approaches to model repair that do not only restore intra-model but also inter-model consistency, i.e., that propagate changes on one kind of model to related models of other kinds. 
Recent such approaches, in particular, with support for synchronising class and sequence diagrams, include~\cite{OhrndorfPKGK21,KretschmerKLE21,BarrigaHRI22,MarchezanKARE23}. 
In these cases, consistency is defined via a modelling language like UML or the requirements of a modelling framework like the Eclipse Modeling Framework~\cite{EMF}, possibly complemented by constraints expressed in a language like OCL~\cite{OMG14}. 
Generally, TGGs could also be applied for such repair tasks. 
For this, it would first be necessary to develop grammars that express the desired consistency relationships between the different kinds of models. 
However, not every consistency requirement expressible in OCL can be expressed using a TGG since for TGGs the language inclusion problem is decidable. 
Having their formal foundation in the theory of graph transformation, TGG-based approaches, including ours, often come with far more comprehensive formal guarantees than offered in approaches like~\cite{OhrndorfPKGK21,KretschmerKLE21,BarrigaHRI22,MarchezanKARE23}. 
On the other hand, whereas TGG-based approaches usually use repair operations that are derived from the rules of the grammar, approaches to model repair are often more flexible. 
For instance, Barriga et al. employ machine learning~\cite{BarrigaHRI22} and Ohrndorf et al.~\cite{OhrndorfPKGK21} and Kretschmer et al.~\cite{KretschmerKLE21} use information on the edit history of a model.
Furthermore, Ohrndorf et al.~\cite{OhrndorfPKGK21} as well as Marchezan et al.~\cite{MarchezanKARE23} do not automatically repair a model but provide the user with a set of (relevant) repair options from which they can choose; in this way, it is guaranteed that user intent will be met. 
Our approach could be extended in this direction by not computing a single \textit{higher-order} \shortcutRule{} but different interesting ones, letting the user select between the different outcomes. 

\section{Conclusion}
\label{chapter:conclusion}
In this paper, we presented a novel approach to construct non-trivial repair rules in the form of higher-order \shortcutRules{} from a given consistency specification.
These repair rules are constructed on demand by analysing a precedence graph that is annotated based on user changes.
Our approach was fully implemented into eMoflon, a state-of-the-art graph transformation tool, which forms the base of our evaluation.
In the evaluation, we showed that higher-order \shortcutRules{} can preserve information in more complex cases, while introducing only a small overhead to the runtime and at times even outperforming other strategies in eMoflon.
Next, we would like to investigate whether we can identify situations statically where a higher-order \shortcutRule{} is needed to improve performance when smaller \shortcutRules{} suffice.
For concurrent synchronisation scenarios, these rules could also be used to check whether sequences of user changes to both models correspond to or contradict each other. 

 \bibliographystyle{alphaurl}
 \bibliography{bibliography}

\appendix
\section{}
\label{chapter:appendix}
This is the entire ILP formulation of our example from \cref{chapter:contribution_extended}.
Note that we added variables for correspondence links, where $co_i$ stands for the correspondence link in between the model elements with the same index $i$.
The correspondence link from \nookRule{} is $nco_i^r$ and the ones from \cubeRule{} are denoted $co_i^r$ and $nco_i^r$, where the first represents the context and the other the created correspondence link.
The correspondence links in \doubleCubeRule{} are named similar to \cubeRule{} but the second created correspondence link connecting \house{} $h_2$ with the \construction{} is called $nnco_i^r$.
Note that all weights are equal 1.
\paragraph{Objective Function}
\begin{gather*}
    max(  v[N_1] + v[C_2] + v[C_3] + v[DC_{1,2}] + \\
     v[p_2^r \rightarrow p_1^r] + v[p_3^r \rightarrow p_1^r] + v[p_{2,3}^r \rightarrow p_1^r] + \\
     v[c_2^r \rightarrow nc_1^r] + v[c_3^r \rightarrow nc_1^r] + v[c_{2,3}^r \rightarrow nc_1^r] + \\
     v[c_3^r \rightarrow nc_1^r] + v[e_2^r \rightarrow e_1^r] + v[e_4^r \rightarrow e_1^r] + \\
     v[e_6^r \rightarrow e_1^r] + v[e_4^r \rightarrow e_3^r] + v[co_2^r \rightarrow nco_1^r] + \\
     v[co_3^r \rightarrow nco_1^r] + v[co_{2,3}^r \rightarrow nco_1^r] + v[co_3^r \rightarrow nco_2^r] + \\ 
     v[p_1^r \rightarrow p_2] + v[nc_1^r \rightarrow c_2] + v[nc_1^r \rightarrow c_3] + \\
     v[nc_2^r \rightarrow c_2] + v[nc_2^r \rightarrow c_3] + v[nc_3^r \rightarrow c_2] + \\
     v[nc_3^r \rightarrow c_3] + v[e_1^r \rightarrow e_1] + v[e_1^r \rightarrow e_2] + \\
     v[e_3^r \rightarrow e_1] + v[e_3^r \rightarrow e_2] + v[e_5^r \rightarrow e_1] + \\
     v[e_5^r \rightarrow e_2] + v[nco_1^r \rightarrow co_2] + v[nco_2^r \rightarrow co_2] + \\
     v[nco_3^r \rightarrow co_2] + v[nco_{1,2}^r \rightarrow co_2] + v[nnco_{1,2}^r \rightarrow co_2] + \\
     v[nco_1^r \rightarrow co_3] + v[nco_2^r \rightarrow co_3] + v[nco_3^r \rightarrow co_3] + \\
     v[nco_{1,2}^r \rightarrow co_3] + v[nnco_{1,2}^r \rightarrow co_3]) \\
\end{gather*}

\paragraph{Rule Candidate Constraints}
\begin{gather*}
    v[C_2] \leq v[N_1] \\
    v[C_3] \leq v[DC_{1,2}] \\
    v[DC_{1,2}] \leq v[N_1] \\
    v[C_2] + v[DC_{1,2}] \leq 1 \\
    v[C_3] + v[DC_{1,2}] \leq 1 \\
\end{gather*}

\paragraph{Context-Create Mapping Constraints}
\begin{align*}
     v[p_2^r \rightarrow p_1^r] &\leq v[N_1] & v[p_2^r \rightarrow p_1^r] &\leq v[C_2] \\ 
     v[p_3^r \rightarrow p_1^r] &\leq v[N_1] & v[p_3^r \rightarrow p_1^r] &\leq v[C_3] \\ 
     v[p_{2,3}^r \rightarrow p_1^r] &\leq v[N_1] & v[p_{2,3}^r \rightarrow p_1^r] &\leq v[DC_{1,2}] \\ 
     v[c_2^r \rightarrow nc_1^r] &\leq v[N_1] & v[c_2^r \rightarrow nc_1^r] &\leq v[C_2] \\
     v[c_3^r \rightarrow nc_1^r] &\leq v[N_1] & v[c_3^r \rightarrow nc_1^r] &\leq v[C_3] \\
     v[c_{2,3}^r \rightarrow nc_1^r] &\leq v[N_1] & v[c_{2,3}^r \rightarrow nc_1^r] &\leq v[DC_{1,2}] \\
     v[c_3^r \rightarrow nc_1^r] &\leq v[N_1] & v[c_3^r \rightarrow nc_1^r] &\leq v[C_3] \\
     v[e_2^r \rightarrow e_1^r] &\leq v[N_1] & v[e_2^r \rightarrow e_1^r] &\leq v[C_2] \\
     v[e_4^r \rightarrow e_1^r] &\leq v[N_1] & v[e_4^r \rightarrow e_1^r] &\leq v[C_3] \\
     v[e_6^r \rightarrow e_1^r] &\leq v[N_1] & v[e_6^r \rightarrow e_1^r] &\leq v[DC_{1,2}] \\
     v[e_4^r \rightarrow e_3^r] &\leq v[C_2] & v[e_4^r \rightarrow e_3^r] &\leq v[C_3] \\
     v[co_2^r \rightarrow nco_1^r] &\leq v[N_1] & v[co_2^r \rightarrow nco_1^r]  &\leq v[C_2] \\
     v[co_3^r \rightarrow nco_1^r] &\leq v[N_1] & v[co_3^r \rightarrow nco_1^r] &\leq v[C_3] \\
     v[co_{2,3}^r \rightarrow nco_1^r] &\leq v[N_1] & v[co_{2,3}^r \rightarrow nco_1^r] &\leq v[DC_{1,2}] \\
     v[co_3^r \rightarrow nco_2^r] &\leq v[C_2] & v[co_3^r \rightarrow nco_2^r] &\leq v[C_3] 
\end{align*}
\begin{align*}
    v[c_3^r \rightarrow nc_1^r] + v[c_2^r \rightarrow nc_2^r] \leq 1 \\
    v[e_4^r \rightarrow e_1^r] + v[e_4^r \rightarrow e_3^r] \leq 1 \\
    v[co_3^r \rightarrow nco_1^r] + v[co_3^r \rightarrow nc_2^r] \leq 1
\end{align*}
\begin{align*}
     v[e_2^r \rightarrow e_1^r] &\leq v[c_2^r \rightarrow nc_1^r] & v[e_2^r \rightarrow e_1^r] &\leq v[p_2^r \rightarrow p_1^r] \\
     v[e_4^r \rightarrow e_1^r] &\leq v[c_3^r \rightarrow nc_1^r] & v[e_4^r \rightarrow e_1^r] &\leq v[p_3^r \rightarrow p_1^r] \\
     v[e_6^r \rightarrow e_1^r] &\leq v[c_{2,3}^r \rightarrow nc_1^r] & v[e_6^r \rightarrow e_1^r] &\leq v[p_{1,2}^r \rightarrow p_1^r] \\
     v[e_4^r \rightarrow e_3^r] &\leq v[c_3^r \rightarrow nc_2^r] & v[e_4^r \rightarrow e_3^r] &\leq v[p_{1,2}^r \rightarrow p_1^r] \\
     v[co_2^r \rightarrow nco_1^r] &\leq v[c_3^r \rightarrow nc_1^r] & v[co_3^r \rightarrow nco_1^r] &\leq v[c_3^r \rightarrow nc_1^r] \\
     v[co_{2,3}^r \rightarrow nco_1^r] &\leq v[c_{2,3}^r \rightarrow nc_1^r]
\end{align*}
\paragraph{Rule-Model Mapping Constraints}
\begin{align*}
     v[p_1^r \rightarrow p_2] &\leq v[N_1] & v[nc_1^r \rightarrow c_2] &\leq v[N_1] \\
     v[nc_1^r \rightarrow c_3] &\leq v[N_1] & v[nc_2^r \rightarrow c_2] &\leq v[C_2] \\
     v[nc_2^r \rightarrow c_3] &\leq v[C_2] & v[nc_3^r \rightarrow c_2] &\leq v[C_3] \\
     v[nc_3^r \rightarrow c_3] &\leq v[C_3] & v[e_1^r \rightarrow e_1] &\leq v[N_1] \\
     v[e_1^r \rightarrow e_2] &\leq v[N_1] & v[e_3^r \rightarrow e_1] &\leq v[C_2] \\
     v[e_3^r \rightarrow e_2] &\leq v[C_2] & v[e_5^r \rightarrow e_1] &\leq v[C_3] \\
     v[e_5^r \rightarrow e_2] &\leq v[C_3] & v[nco_1^r \rightarrow co_2] &\leq v[N_1] \\
     v[nco_2^r \rightarrow co_2] &\leq v[C_2] & v[nco_3^r \rightarrow co_2] &\leq v[C_3] \\
     v[nco_{1,2}^r \rightarrow co_2] &\leq v[DC_{1,2}] & v[nco_{1,2}^r \rightarrow co_2] &\leq v[DC_{1,2}]\\
     v[nco_1^r \rightarrow co_3] &\leq v[N_1] & v[nco_2^r \rightarrow co_3] &\leq v[C_2] \\
     v[nco_3^r \rightarrow co_3] &\leq v[C_3] & v[nco_{1,2}^r \rightarrow co_3] &\leq v[DC_{1,2}] \\
     v[nnco_{1,2}^r \rightarrow co_3]] &\leq v[DC_{1,2}]
\end{align*}
\begin{align*}
     v[nc_1^r \rightarrow c_2] + v[nc_1^r \rightarrow c_3] \leq 1 \\
     v[nc_2^r \rightarrow c_2] + v[nc_2^r \rightarrow c_3] \leq 1 \\
     v[nc_3^r \rightarrow c_2] + v[nc_3^r \rightarrow c_3] \leq 1 \\
     v[e_1^r \rightarrow e_1] + v[e_1^r \rightarrow e_2] \leq 1 \\
     v[e_3^r \rightarrow e_1] + v[e_3^r \rightarrow e_2] \leq 1 \\
     v[e_5^r \rightarrow e_1] + v[e_5^r \rightarrow e_2] \leq 1 \\
     v[nco_1^r \rightarrow co_2] + v[nco_1^r \rightarrow co_3] \leq 1 \\
     v[nco_2^r \rightarrow co_2] + v[nco_2^r \rightarrow co_3] \leq 1 \\
     v[nco_3^r \rightarrow co_2] + v[nco_3^r \rightarrow co_3] \leq 1 \\
     v[nco_{1,2}^r \rightarrow co_2] + v[nco_{1,2}^r \rightarrow co_3] \leq 1 \\
     v[nnco_{1,2}^r \rightarrow co_2] + v[nnco_{1,2}^r \rightarrow co_3] \leq 1
\end{align*}
\begin{align*}
    3 v[e_3^r \rightarrow e_1] &= v[nc_2^r \rightarrow c_2] + v[p_1^r \rightarrow p_2] + v[p_2^r \rightarrow p_1^r] \\
    3 v[e_3^r \rightarrow e_2] &= v[nc_2^r \rightarrow c_3] + v[p_1^r \rightarrow p_2] + v[p_2^r \rightarrow p_1^r] \\
    3 v[e_5^r \rightarrow e_1] &= v[nc_3^r \rightarrow c_2] + v[p_1^r \rightarrow p_2] + v[p_3^r \rightarrow p_1^r] \\
    3 v[e_5^r \rightarrow e_2] &= v[nc_3^r \rightarrow c_3] + v[p_1^r \rightarrow p_2] + v[p_3^r \rightarrow p_1^r] \\
\end{align*}
\begin{align*}    
    v[e_1^r \rightarrow e_1] &\leq v[nc_1^r \rightarrow c_2] & v[e_1^r \rightarrow e_1] &\leq v[p_1^r \rightarrow p_2] \\
    v[e_1^r \rightarrow e_2] &\leq v[nc_1^r \rightarrow c_3] & v[e_1^r \rightarrow e_2] &\leq v[p_1^r \rightarrow p_2] \\
    v[nco_2^r \rightarrow co_2] &\leq v[nc_2^r \rightarrow c_2] & v[nco_3^r \rightarrow co_3] &\leq v[nc_3^r \rightarrow c_3] \\
    v[nco_{1,2}^r \rightarrow co_2] &\leq v[nc_3^r \rightarrow c_2] & v[nnco_{1,2}^r \rightarrow co_3] &\leq v[nc_3^r \rightarrow c_3]\\
\end{align*}

\end{document}